\documentclass{aa}  

\usepackage{graphicx}
\usepackage{txfonts}
\usepackage[]{hyperref}
\usepackage{amsmath}	
\usepackage{amssymb}	
\usepackage{xcolor}		
\usepackage{breakcites} 
\usepackage{caption} 
\usepackage{placeins}   
\usepackage[export]{adjustbox} 

\usepackage{cellspace}  
\newcommand{  \Halpha   }{\ifmmode {\rm H}\alpha \else H$\alpha$\fi}

\newcommand{  \Hbeta    }{\ifmmode {\rm H}\beta \else H$\beta$\fi}

\newcommand{  \Hgamma   }{\ifmmode {\rm H}\gamma \else H$\gamma$\fi}
\newcommand{  \Hdelta   }{\ifmmode {\rm H}\delta \else H$\delta$\fi}
\newcommand{  \Lya      }{\ifmmode {\rm Ly}\alpha \else Ly$\alpha$\fi}
\newcommand{\oii}{\ifmmode \left[{\rm O}\,\textsc{ii}\right] \else [O\,{\sc ii}]\fi}
\newcommand{\OII}{\ifmmode \left[{\rm O}\,\textsc{ii}\right]\,\lambda 3726, 3729 \else [O\,{\sc ii}]\,$\lambda 3726, 3729$\fi}
\newcommand{\oiii}{\ifmmode \left[{\rm O}\,\textsc{iii}\right] \else [O\,{\sc iii}]\fi}
\newcommand{\OIII}{\ifmmode \left[{\rm O}\,\textsc{iii}\right]\,\lambda5007 \else [O\,{\sc iii}]\,$\lambda5007$\fi}
\newcommand{  \nii      }{\ifmmode \left[{\rm N}\,\textsc{ii}\right]  \else [N\,\textsc{ii}]\fi}
\newcommand{  \sii      }{\ifmmode {\rm \left[S}\,\textsc{ii} \right] \else [S\,\textsc{ii}]\fi}
\newcommand{\SII}{\ifmmode \left[{\rm S}\,\textsc{ii}\right]\,\lambda6716, 6731 \else [S\,{\sc ii}]\,$\lambda 6716, 6731$\fi}
\newcommand{\Msun}{\ifmmode M_{\odot} \else $M_{\odot}$\fi}
\newcommand{\Lsun}{\ifmmode L_{\odot} \else $L_{\odot}$\fi}
\newcommand{\Zsun}{\ifmmode Z_{\odot} \else $Z_{\odot}$\fi}

\begin{document}

   \title{
    A cosmic collision in the making: JWST/NIRISS reveals a merging and maturing protocluster core at z $\sim $ 3 
    around a red quasar
   }
    \titlerunning{A merging and maturing protocluster at $z=3$}

   \author{C. Bertemes\inst{1}
          \and D. Wylezalek\inst{1} 
          \and G. Noirot\inst{2}
          \and R. Hviding\inst{3}
        \and D. S. N. Rupke\inst{4,1}
        \and N. L. Zakamska\inst{5,6}
        \and S. Veilleux\inst{7}
        \and L. Gatto\inst{8,1}
        \and W. Liu\inst{9}
        \and A. Vayner\inst{10}
        \and Y. Ishikawa\inst{11}
        \and Y. Chen\inst{5}
        \and S. Sankar\inst{5}
          }

   \institute{Zentrum f\"ur Astronomie der Universit\"at Heidelberg, Astronomisches Rechen-Institut M\"onchhofstr, 12-14 69120 Heidelberg, Germany\\
              \email{c.bertemes@uni-heidelberg.de}
   \and 
   Space Telescope Science Institute, 3700 San Martin Drive, Baltimore, MD 21218, USA
   \and
   Max-Planck-Institut für Astronomie, Königstuhl 17, D-69117 Heidelberg, Germany
    \and
    Department of Physics, Rhodes College, Memphis, TN 38112, USA
   \and
    Department of Physics and Astronomy, Bloomberg Center, Johns Hopkins University, Baltimore, MD 21218, USA
    \and 
    Institute for Advanced Study, Princeton, NJ 08540, USA
    \and 
    Department of Astronomy and Joint Space-Science Institute, University of Maryland, College Park, MD 20742, USA
    \and 
    Departamento de Astronomia, Instituto de Física, Universidade Federal do Rio Grande do Sul, Av. Bento Gonçalves 9500, Porto Alegre, RS, 91501-970, Brazil
    \and 
    Steward Observatory, University of Arizona, 933 N. Cherry Ave., Tucson, AZ 85721, USA
    \and
    Florida Gulf Coast University, 10501 FGCU Blvd. South, Fort Myers, 33965, FL, USA
    \and
    MIT Kavli Institute for Astrophysics and Space Research, Massachusetts Institute of Technology, Cambridge, MA 02139, USA
    }
   \date{Received XXX; accepted YYY}

\abstract
{We report the discovery of ``The Step'', a dense protocluster at $z = 3$ revealed by JWST/NIRISS Wide-Field Slitless Spectroscopy (WFSS) of the field around the luminous quasar and starburst SDSSJ165202.64+172852.3. The protocluster contains at least $18$ member galaxies and exhibits signs for two substructures in the 3D position-velocity space -- notably a distinctive step-like distribution, from which the nickname derives. 
We propose that the Step may be undergoing a merger of two haloes, which could also explain the unusually large velocity span ($700-850$km/s) between galaxies along the densest line of sight. This sightline contains four/seven galaxies within projected distances of $20$/$70$ kpc {from the quasar}. {Within a $320 \times 230$ kpc$^2$ region, we find $10$ galaxies (including the quasar), consistent with a remarkably dense protocluster core. } 
We derive a galaxy overdensity {of a factor $>100$ in this core and $>15$ in the full field (1 Mpc$^2$),} making it one of the densest structures known at this redshift, and a very massive halo of $13.1 < \log  M_{\rm halo}/ M_\odot < 13.8$, which suggests the structure could potentially evolve into a massive, Coma-like cluster at $z=0$. Compared to the field, the star-forming galaxies in the Step appear to show slightly elevated levels of star formation activity. Yet, we find two quiescent galaxies ($\log M_\star / M_\odot \sim 10.2, 10.4 \pm 0.8$) with strong continuum breaks and no evidence for ongoing star formation. Compared to other protoclusters at $z=3$, the Step structure exhibits broadly similar, or slightly suppressed star-forming activity for its halo mass. We also identify two new AGN candidates based on optical line ratios and find tentative evidence for an enhanced AGN fraction ($10-20 \%$) with respect to the field. The protocluster constitutes one of the emerging handful of maturing structures at $2 \leqslant z \leqslant 4$ that host both star-forming and quenched galaxies. This redshift regime may thus be a key transitional epoch for cluster studies, where overdensities transition from being the most active sites of star formation in the early Universe towards shaping the massive passive ellipticals seen in the cores of today's clusters. 
}


   \maketitle

\section{Introduction}
\label{sec:intro}

Galaxies in overdense regions are known to evolve and quench more rapidly than field galaxies due to environmental processes \citep{Dressler1980}, which include ram-pressure stripping, strangulation, and more frequent galaxy-galaxy interactions \citep{Boselli2006}. Clusters at low redshifts are dense, virialised, sub-Mpc regions typically dominated by massive, passive ellipticals at their core \citep{Mei2009}, with little ongoing star formation. Their evolved state compared to the field implies that their progenitors - non-virialised protoclusters at high redshift - must have experienced a period of accelerated growth at early cosmic times. The timing and duration of the turnover in star-forming activity remain an active area of research in cluster studies. 

Semi-analytic models suggest that protoclusters contribute significantly to the cosmic star formation rate (SFR) density, by over $20$\% at redshifts $z > 2$ \citep{Chiang2017}, with the contribution rising as a function of redshift. 
Observational studies of high-redshift protoclusters have built up gradually over time, given the challenge of finding such systems in the first place. 
For instance, based on 16 overdense structures at $1.4 < z < 2.8$, \citet{Noirot2018} found that star formation in member galaxies is increasing with redshift up to $z=2$, though the most massive galaxies at the core show signs of suppressed star formation at all redshifts. At $z \sim 2.5-4$, the majority of protocluster observations point to enhanced or similar levels of star formation compared to the field \citep{Alexander2016, Casey2016, Kubo2019, Alberts2022} as well as {increased or comparable molecular gas reservoirs} \citep{Polletta2022, Pensabene2024, Zhou2025}. However, recent observations have brought to light a handful of protocluster systems hosting one or more massive quiescent galaxies (\citealt{Kubo2021}, \citealt{Shi2021}, \citealt{McConachie2022}, \citealt{Ito2023}, \citealt{Naufal2024}, \citealt{TanakaM2024}, {\citealt{Kakimoto2026}}). These systems may have eluded previous detection partly because protocluster identification often relies on using star-forming galaxies as tracers (see e.g. the review by \citealt{Overzier2016}), and partly due to sensitivity limits that prevent the detection of faint, line-less passive galaxies. 

In recent years, the James Webb Space Telescope (JWST) has unveiled an unexpectedly high density of massive quiescent galaxy candidates at high redshift \citep{Carnall2023}, and the role of environment in their quenching remains poorly understood. \citet{Helton2024} conducted a systematic search of $z>4.9$ overdensities in the GOODS fields and found that the brightest, reddest objects reside in denser environments. This trend is consistent with cosmological simulations, which predict that quenched galaxies at high redshift preferentially inhabit overdense regions, where they experience enhanced feedback from Active Galactic Nuclei (AGN) \citep{KurinchiVendhan2024}. {Recent observations of the SpiderWeb protocluster at $z=2.2$ indeed uncovered a red sequence of galaxies with a high (mostly X-Ray) AGN fraction of up to $\sim 50\%$ \citep{Tozzi2022, Naufal2024}.} {Further, radio-loud AGN have been used as signposts for identifying protoclusters at high redshift (e.g. \citealt{Galametz2010, Wylezalek2013, Noirot2016, PaternoMahler2017, Noirot2018}).} It has also been proposed that AGN are triggered by mergers in dense environments \citep{Hopkins2008}, although the evidence for a link between AGN and merger signatures remains mixed \citep{Kocevski2012, Glikman2015, Kocevski2015, Villforth2017, Wylezalek2016}. Beyond redshift $z \sim 2$, studies characterising the incidence of AGN as a function of large-scale environment are still scarce, with contradicting results -- either supporting a higher incidence of quasars in protoclusters \citep{Lehmer2009, Digby-North2010} or not \citep{Macuga2019}.

When the JWST Early Release Science program Q3D targeted  the luminous red quasar SDSSJ165202.64+172852.3 (dubbed J1652) at $z_{\rm spec}=2.9489$ with the Near-InfraRed Spectrograph (NIRSpec) IFU, the spatially resolved spectra unexpectedly revealed three galaxies within projected distances of $<20$kpc with a very high velocity range of $700$ km/s between each other, as well as an extended tidal tail which is also detected in the stellar continuum \citep{Wylezalek2022b}. 
Studies of the extended \oiii\ kinematics reveal both turbulence signatures driven by the quasar on scales of $<10$ kpc, but also an excess of turbulence in the outer nebula (10-20 kpc), suggesting an additional source of energy injection which may be linked to the environment \citep{ChenM2025}. 
Imaging from the Hubble Space Telescope also shows a clear overdense structure westwards of J1652 \citep{Zakamska2019}. 
The galaxy is at a mature stage, with a stellar mass of $ \log M_\star /  M_\odot \sim 10.9$ \citep{Chen2026}, exceeding the characteristic mass of the $z=3$ stellar mass function \citep{Weaver2023}, and a black hole weighing $\gtrsim 10^9$ solar masses ({\citealt{Perrotta2019, Bertemes2025}}). The quasar contains starbursting clumps forming $\geqslant 200 \ M_\odot$/yr, as well as shock-ionised regions and powerful extended winds exceeding $2000$ km/s \citep{Alexandroff2018, Vayner2023, Vayner2024}. The complex morphology, the dense population of galaxies and the substantial velocity spread between the companions ($\sim 700$ km/s) suggest that this system may represent the core of a forming protocluster. J1652 further belongs to a peculiar class of AGN called `extremely red quasars' \citep{Hamann2017}. These sources are characterised by UV-to-IR colours that are orders of magnitude redder than blue quasars, powerful outflows among the fastest known to date (\citealt{Zakamska2016, perr19}), and very massive host galaxies \citep{zaka19}. Given that the most massive galaxies, to first order, live in the most massive halos, extremely red quasars may likewise be expected to reside in overdensities, which could potentially contribute to triggering this particular quasar phase. 

With new JWST follow-up Wide-Field Slitless Spectroscopy (WFSS) observations using the NIRISS instrument {(Program ID: GO 3807; PI: Bertemes)}, we obtain slitless spectroscopy for the wide field ($2.2 \times 2.2$ arcmin) around J1652 to test whether the quasar indeed resides in a protocluster. We use two filters centred on $1.5$ and $2 \mu$m, which cover the \oiii , \Hbeta\ and \oii lines, as well as the $4000 \AA$ break and Balmer break at the redshift of the quasar. Besides redshift identification, our observations yield insights into the overdensity of the field, the fraction of passive galaxies and AGN, as well as the levels of star formation. The observed kinematics enable us to constrain the halo mass and provide evidence that the structure may be undergoing a merger of two haloes. 

The paper is organised as follows: {In Section \ref{sec:obs_methods}, we describe the observations, the data reduction, as well as the redshift fitting.} In Section 3, we discuss the protocluster's structure, halo mass, as well as properties of its member galaxies, and we put our results into context with the literature. Section 4 summarises our conclusions.

Throughout the paper, we use a we use a $\Lambda$CDM cosmology defined by $\Omega_{\Lambda} = 0.7$, $\Omega_{\rm m} = 0.3$, and $H_0 = 70\ {\rm km}\ {\rm s}^{-1}\ {\rm Mpc}^{-1}$. We use proper distances {unless otherwise stated}. At the redshift of the target, one arcsec corresponds to $7.74$ kpc.

\section{Observations and methods}
\label{sec:obs_methods}

\subsection{JWST observations}
\label{sec:obs}

The field around J1652 was observed with JWST/NIRISS on July 2nd 2024 using an 8-point MEDIUM dithering pattern. The total exposure time was $2.67$h, with $1.72$h for grism observations with the F150W filter, $0.76$h for the F200W filter, as well as $344$s of photometric images in each filter for calibration purposes. The grism exposure time was divided equally between the two orthogonal grisms, GR150R and GR150C. 

\subsection{Data reduction}
\label{sec:data_reduction}

We obtained the raw JWST/NIRISS data from the Mikulski Archive for Space Telescopes (MAST) and reduced it with version 1.16.1 of the JWST pipeline\footnote{\url{https://github.com/spacetelescope/jwst}}. We performed detector-level calibrations on the individual exposures using the \texttt{calwebb\_detector1} pipeline with version 12.0.6 and context 1321 of the Calibration Reference Data System (CRDS). We removed random jumps along the detector columns via the \texttt{ColumnJump} step \footnote{\url{https://github.com/chriswillott/jwst}} using a $5 \sigma$ threshold both for flagging a single or two jumps, before running the \texttt{jump} step. 
For the direct images, we flat-fielded the data prematurely after the \texttt{jump} step (which was performed with a rejection threshold of $5 \sigma$), conducted a 1/f correction at the group level via the \texttt{clean\_flicker\_noise} step with the `median' background method\footnote{following the \texttt{image1overf} routine: \url{https://github.com/chriswillott/jwst}} and masking parameters \texttt{sigma\_bgmask}, \texttt{sigma\_1fmask} $= 3, 2$. We re-applied the flat-field again before proceeding with the rest of the pipeline steps.

With stage 1 of the JWST data reduction pipeline complete, we performed a 1/f removal on the individual rate files via \texttt{image1overf} for both the WFSS exposures and the direct images. For the latter, this was done in addition to the aforementioned 1/f correction at the group level, as this yielded further improvement. 
We proceeded with the \texttt{grizli} software \citep{Brammer_grizli}, which provides its own implementation of typical stage 2 steps such as flat-fielding, WCS assignment, and photometric scaling. We background-subtracted both the WFSS files and the direct images using the empirical backgrounds by \citet{Hviding2024}. The latter have been shown to agree with the updated empirical CRDS backgrounds as of context 1365 presented in Noirot et al. (\citeyear{Noirot2025})  down to the $1\%$ level. We modelled the location of 2D spectra and contamination via overlapping traces based on the positions and morphologies of sources in the direct image using two iterations and linear trace models within the wavelength range of a given filter (with initial coefficients [1, -0.6]).  The contamination was modelled for sources down to a limiting magnitude \texttt{prelim\_mag\_limit} $=26$. and neglected for fainter sources.  We used the 221215 configuration files by \citet{Matharu_221215}, who empirically determined the pixel-dependent trace location and wavelength zeropoint. The source catalogue from the direct images was generated in \texttt{grizli} via the \texttt{SEP} python package \citep{Barbary2016} wrapping around the \texttt{SOURCE EXTRACTOR} tool \citet{Bertin1996}, using a detection threshold of $1.2 \sigma$, yielding $1498$ sources. Finally, the 1D spectra are obtained by summing the traces along the non-spectral axis.

\subsection{Redshift identification and spectral fitting}
\label{sec:redshifts}

{Within the $2.2 \times 2.2$ arcmin$^2$ ($\sim 1$ Mpc$^2$ at $z=3$) NIRISS footprint centred on J1652, we conduct a blind search for galaxies at a similar redshift to the quasar. to this end,} we use \texttt{grizli} for redshift identification by fitting the spectra with scaled-up templates from the Flexible Stellar Population Synthesis (FSPS) library \citep{Conroy2009, Conroy2010}. Based on the source morphology in the direct image, \texttt{grizli} forward-models the 2D spectra of the templates in the grism frames. We used a redshift range of $0-10$, and the `nnls' fitter (i.e., non-negative least squares) and the `bounded' option enforcing non-negative continuum templates but allowing negative line templates. After inspecting all $72$ candidates for which the fitting routine returned an unambiguous redshift in the range [$2.8$, $3.1$], we identify robust protocluster member galaxies via the presence of two spectral features (\oii\ and \oiii\ lines, or a $4000\AA$ break with Ca H and K absorption) seen in the 1D and 2D spectra of both orthogonal grisms. {As all 17 candidates selected in that manner are offset by less than 2000 km/s from the quasar, we consider them all to be part of the protocluster.} We also visually inspect the 2D spectra for contamination and artifacts. 
The resolving power of the grisms is $R \sim 150$ (set by the undersampled NIRISS point spread function). A simple Monte Carlo exercise suggests uncertainties on the recovered redshifts below $\sim0.003$ across our sample
\footnote{We generate Gaussian mock emission lines with an intrinsic velocity dispersion of 500 km/s at varying peak positions. The lines are convolved with the PSF, resampled onto the grism wavelength grid, perturbed within the observed rms noise for each source, and then fitted to compare the recovered redshifts to the true input.}. Calibration uncertainties and systematics may further affect the results. We therefore conservatively assume a uniform error of $0.005$ on the recovered redshifts. {Within these uncertainties, we assume the redshifts to be robust, given our criterion of detecting at least two spectral features.}

The observed redshift of a given target can be related to the quasar's redshift by interpreting the difference either as a spatial separation between the two (i.e., different Hubble flow velocities) or an offset in their {peculiar} velocity along the line of sight, or a combination of both.  
Henceforth, we will use both interpretations interchangeably, since we cannot distinguish between the two.

\setlength{\tabcolsep}{4.5pt}
\begin{table*}
\centering
\caption{List of high-confidence protocluster member galaxies. 
We adopt a uniform redshift uncertainty of $0.005$.}
\label{tab:sample}
\begin{tabular}{ Sl Sr Sr  Sr Sc Sr Sr Sr}
\hline
    ID &        RA  &        DEC &  {Object Name} &  $z$ &      d$_{\rm proj}$ &    d$_{\rm 3D}$ & {Structure} \\
       &      [deg] &      [deg] &    &      &                  [arcsec] &                     [Mpc] & \\
\hline
   J1652 &   253.01100 &    17.48119 & J165202.64+172852.28  &    2.9489 &           &     & Main  \\
     1 &  253.01106 &   17.49700 & J165202.65+172949.21 & $     2.940$ &                     56.81 & <                    2.02 & Subgroup\\
2 ($\rm P_N$) &  253.00862 &   17.49163 & J165202.07+172929.88 & $     2.931$ &                     38.48 & <                    4.34  & Subgroup\\
     3 &  253.00247 &   17.49107 & J165200.59+172927.84 & $     2.943$ &                     46.95 & <                    1.25  & Subgroup\\
     4 &  252.99066 &   17.48528 & J165157.76+172907.00 & $     2.923$ &                     74.73 & <                    6.32  & Subgroup\\
5 ($\rm LR_{1}$) &  253.03087 &   17.48367 & J165207.41+172901.22 & $     2.932$ &                     72.01 & <                    3.97  & Subgroup\\
6 ($\rm LR_{2}$) &  253.02763 &   17.48219 & J165206.63+172855.88 & $     2.927$ &                     59.88 & <                    5.13  & Subgroup\\
     7 &  253.01157 &   17.48145  & J165202.78+172853.22 & $     2.957$ &                      2.16 & <                    2.33  & Main\\
     8 &  253.01125 &   17.48104  & J165202.70+172851.73 & $     2.956$ &                      1.05 & <                    1.88  & Main\\
     9 &  253.01132 &   17.48092  & J165202.72+172851.32 & $     2.956$ &                      1.51 & <                    1.93  & Main\\
10 ($\rm B_{2}$) &  253.00904 &   17.47996 & J165202.17+172847.85 & $     2.946$ &                      8.45 & <                    0.42  & Main\\
    11 &  253.01112 &   17.47978 & J165202.67+172847.22 & $     2.951$ &                      5.18 & <                    0.77  & Main\\
12 ($\rm B_{1}$) &  253.01111 &   17.47893 & J165202.67+172844.16 & $     2.946$ &                      8.24 & <                    0.41  & Main\\
    13 &  253.01310 &   17.47408 & J165203.14+172826.67 & $     2.943$ &                     26.79 & <                    1.20  & Main\\
    14 &  253.00523 &   17.47376 & J165201.26+172825.54 & $     2.951$ &                     33.99 & <                    0.81  & Main\\
    15 &  253.02534 &   17.47354 & J165206.08+172824.73 & $     2.937$ &                     58.50 & <                    2.80  & Main\\
16 ($\rm P_S$) &  253.00996 &   17.47003 & J165202.39+172812.11 & $     2.946$ &                     40.46 & <                    0.52  & Main\\
    17 &  253.00953 &   17.46313 & J165202.29+172747.26 & $     2.961$ &                     65.35 & <                    3.15  & Main\\
\hline
\end{tabular}
\tablefoot{For each source, we list the RA and DEC, {object name derived from its sexagesimal coordinates,} redshift, the projected distance from the quasar J1652 in arcsec, the deprojected distance if interpreting the observed redshifts as spatial separations based on the Hubble flow (which is an upper limit){, and whether they belong to the main halo or the subgroup identified in Fig \ref{fig:protocluster_field_3D}}. The IDs are assigned in order of decreasing DEC, with alternative designations for noteworthy galaxies: $\rm P_N$/$\rm P_S$ refers to the Northern/Southern passive source, $\rm LR_{1}$ and $\rm LR_{2}$ are the AGN candidates with high \oiii/\Hbeta\ line ratios, and $\rm B_{1}$ and $\rm B_{2}$ are the galaxies with broad \oiii .} 
\end{table*}
\setlength{\tabcolsep}{6pt}

\section{Results}
\label{sec:results}

\subsection{Protocluster member galaxies}
\label{sec:PC_members}

\begin{figure*}[]
{
    \centering
    \includegraphics[width=0.62\linewidth, valign=m]{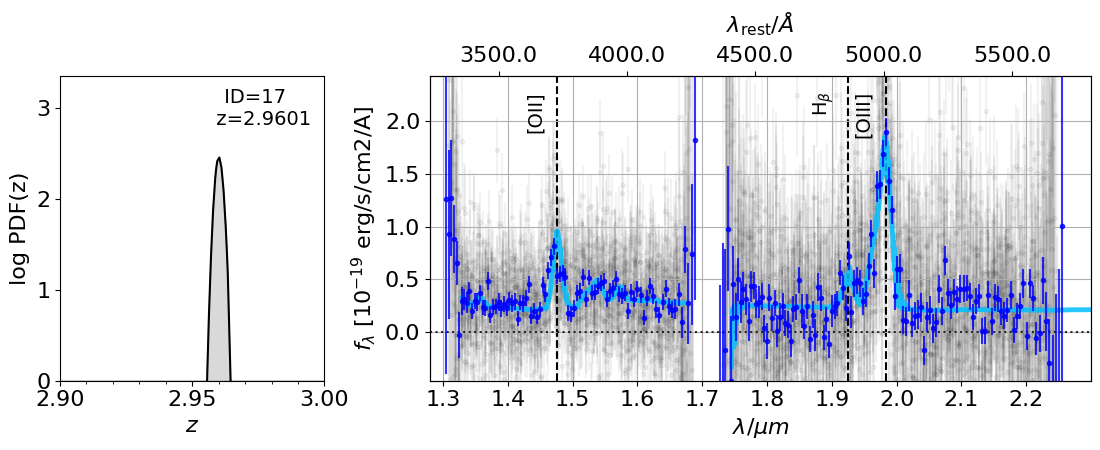}
    \includegraphics[width=0.23\linewidth, valign=m, trim={6cm 1.5cm 6cm 2cm}, clip]{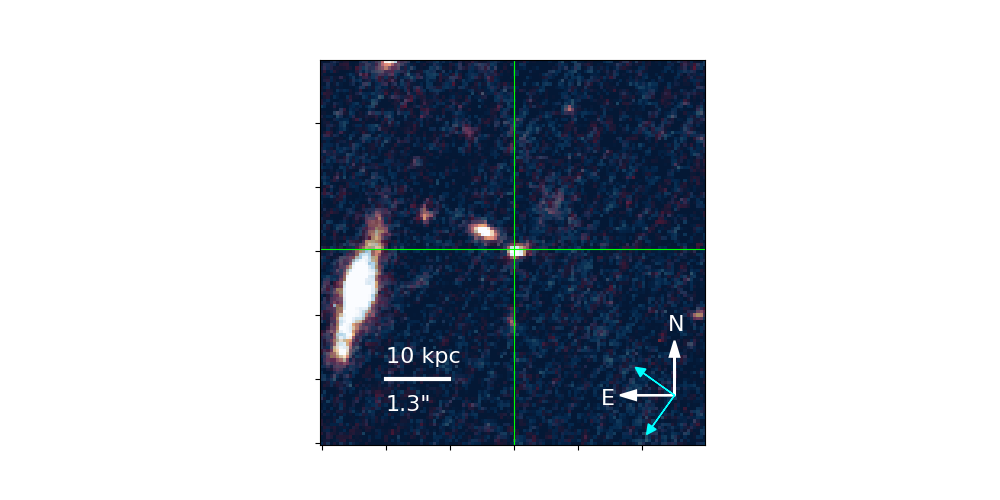}
    \includegraphics[width=0.62\linewidth, valign=m]{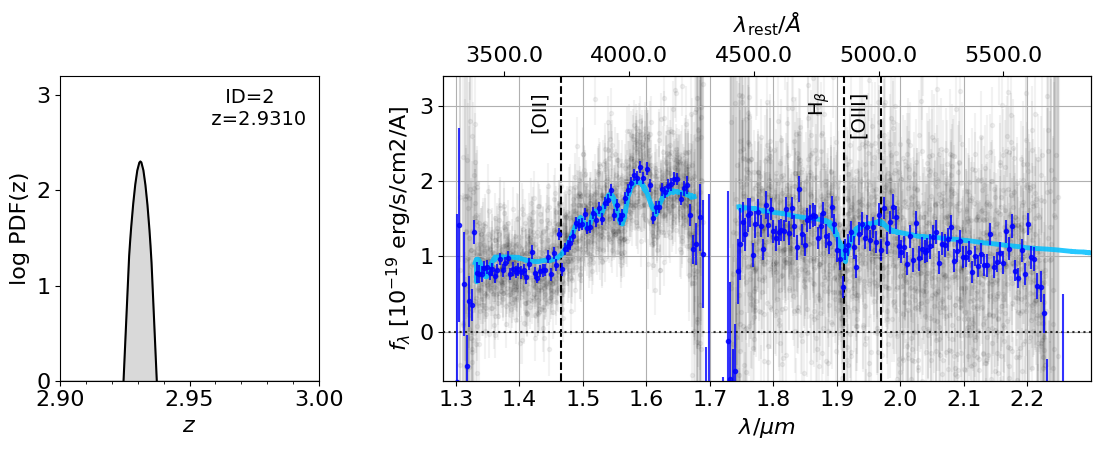}
    \includegraphics[width=0.23\linewidth, valign=m, trim={6cm 1.5cm 6cm 2cm}, clip]{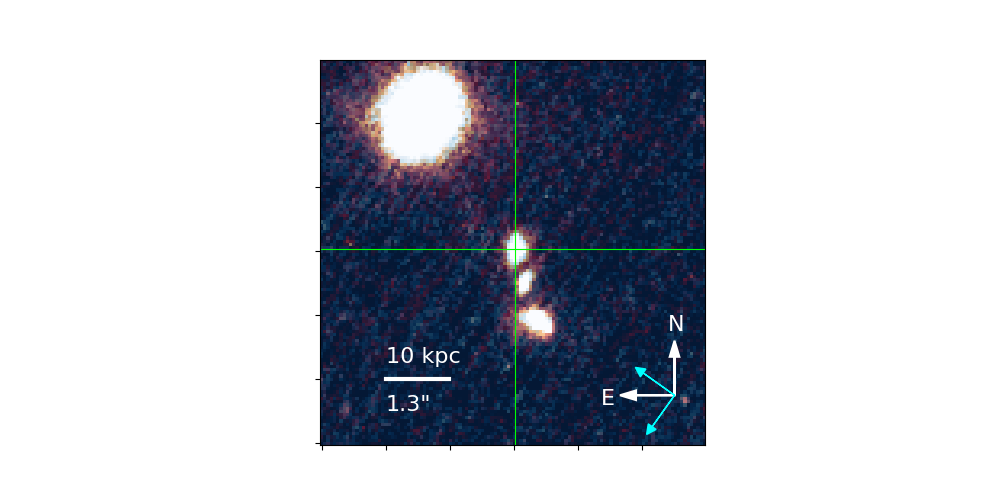}
    \includegraphics[width=0.62\linewidth, valign=m]{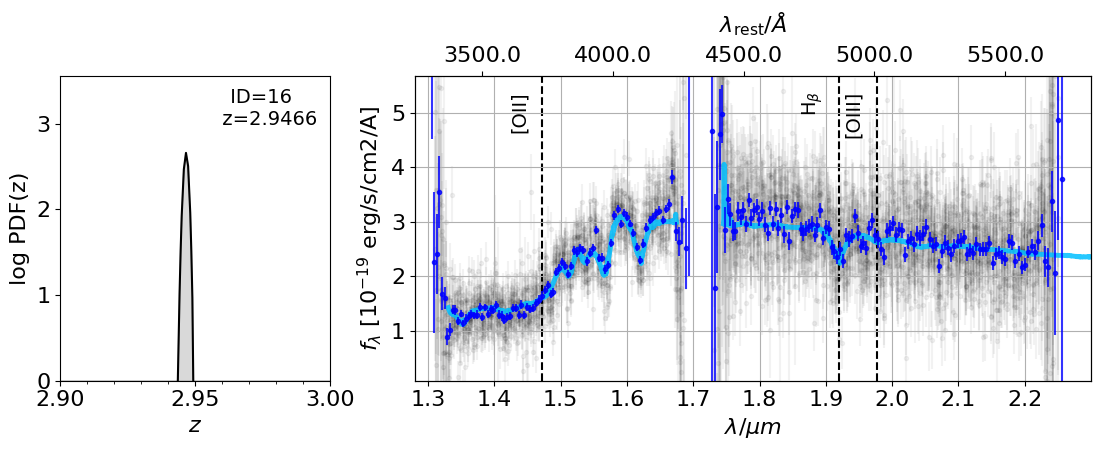}
    \includegraphics[width=0.23\linewidth, valign=m, trim={6cm 1.5cm 6cm 2cm}, clip]{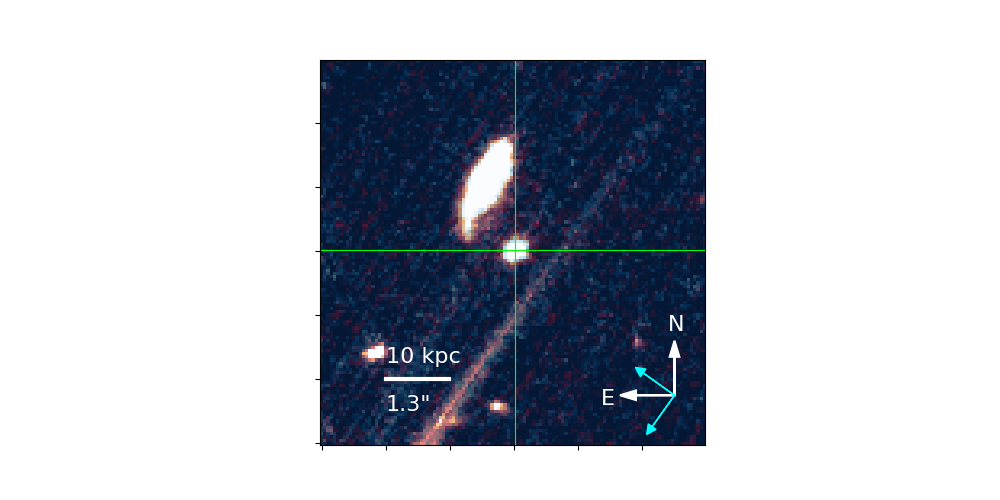}
    \caption{Spectra and recovered redshift probability distribution of a representative star-forming galaxy and the two passive galaxies, as well as cutout composite images based on the HST WFC3 (blue) and ACS (red) photometry. The orientations of the grism dispersers are indicated by cyan arrows. The spectral plots show the datapoints from the individual spectral traces in grey, the 1D spectrum consisting of averages in dark blue, and the best-fit template in light blue. The spectra of the remaining cluster members are presented in Appendix \ref{sec:app} (Fig \ref{fig:other_spectra}). }
    \label{fig:spectra}
}
\end{figure*}

\begin{figure}[]
{
    \begin{minipage}{\linewidth}
        \begin{minipage}{0.92\linewidth}
        \hspace*{1.5cm} F150W \hspace*{2.5cm} F200W  \hspace*{4cm} \phantom{F150W} \hspace*{2.5cm} \phantom{F200W}
        \vspace*{-0.5cm}
        \end{minipage}
    \end{minipage}
    \begin{minipage}{0.125\linewidth}
            Grism 1 \\[3.5ex]
            Grism 2 \\[3.5ex]
            Line      \vspace*{0.7cm}
    \end{minipage}
    \begin{minipage}{0.87\linewidth}
        \includegraphics[width=\linewidth, trim={0cm, 0.8cm, 0cm, 0cm}, clip]{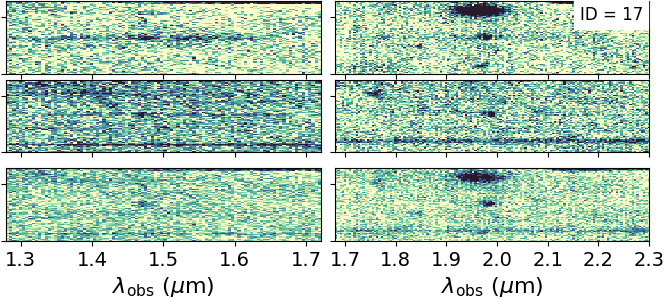}
    \vspace*{0.05cm}
    \end{minipage}
    \begin{minipage}{0.125\linewidth}
            Grism 1 \\[3.5ex]
            Grism 2 \\[3.5ex]
            Line      \vspace*{0.7cm}
    \end{minipage}
    \begin{minipage}{0.87\linewidth}
        \includegraphics[width=\linewidth, trim={0cm, 0.8cm, 0cm, 0cm}, clip]{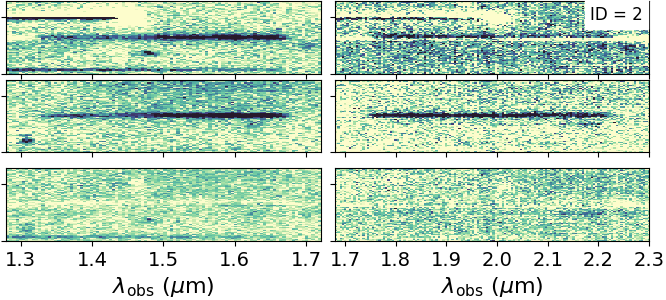}
    \vspace*{0.05cm}
    \end{minipage}
    \begin{minipage}{0.125\linewidth}
            Grism 1 \\[3.5ex]
            Grism 2 \\[3.5ex]
            Line      \vspace*{0.7cm}
    \end{minipage}
    \begin{minipage}{0.87\linewidth}
        \includegraphics[width=\linewidth, trim={0cm, 0cm, 0cm, 0cm}, clip]{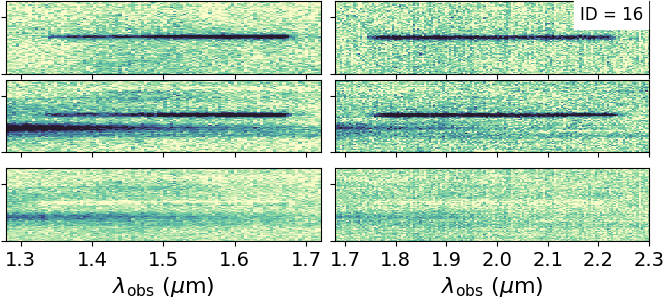}
    \end{minipage}
\caption{2D NIRISS spectra of the representative star-forming protocluster member galaxy and the two line-less passive galaxies from Fig \ref{fig:spectra} (see Fig \ref{fig:2D_spectra_rest} for the remaining  member galaxies). The left and right columns show the spectra in the F150W and F200W filters, respectively, in observed frame. The first and second row show the spectra in both orthogonal grisms separately, while the third row shows the line emission after subtracting the modelled continuum based on a stack of the two grisms (without correcting for position angle). The two passive galaxies show no evidence for line emission.}
\label{fig:2D_spectra}
}
\end{figure}

Following the data reduction and redshift identification procedure outlined in Sections \ref{sec:data_reduction} and \ref{sec:redshifts}, we obtain a sample of $18$ robust protocluster members with redshifts ranging from $2.923$ to $2.961$. This sample includes the quasar J1652 and the three galaxies identified by the NIRSpec IFU data (\citealt{Wylezalek2022b} and Section \ref{sec:intro}), whose slitless spectra by NIRISS are too contaminated to be used. The member galaxies' coordinates and their distances with respect to J1652 are listed in Table \ref{tab:sample}{, and we report their \oii, \oiii\ and \Hbeta\ fluxes in Table \ref{tab:fluxes}}. We also flag one potential additional member galaxy in Appendix \ref{app:pot_extra}, which we however exclude from the analysis due to an ambiguity in redshift due to the low signal-to-noise. Out of the sample, $11$ galaxies lie within a 3D distance of $\lesssim 2$ Mpc from J1652 (if interpreting the observed redshifts as spatial offsets), and all of them fall within a redshift range sufficiently narrow for the structure to clearly meet the definition of a protocluster as given by \citet{Eisenhardt2008} (containing $\geq 5$ members within 2 Mpc and a redshift range $z \pm 0.06*(1+z)$). {We thus confirm the protocluster hypothesis proposed by \citet{Wylezalek2022b}.}

In the top panel of Fig \ref{fig:spectra}, we show an example spectrum of a representative galaxy (ID 17) with a clear \oiii\ detection at a signal-to-noise ${\rm (S/N)_{\oiii} }=23$ typical for our sample which spans ${\rm (S/N)_{\oiii} } = 3-70$. \Hbeta\ is detected at a S/N of $3.8$, and we report \Hbeta\ detections at the $\gtrsim 3 \sigma$ level in 15 out of 18 sources, indicating that the majority of protocluster members are star-forming. The 2D spectrum of ID 17 is shown in Fig \ref{fig:2D_spectra} (top). 

Intriguingly, our slitless spectra reveal two galaxies with a clear $4000$ \AA\ break (D4000) and Balmer break (IDs 2 and 16), with no evidence for ongoing star formation given the absence of \Hbeta\ emission. Their 1D and 2D spectra are also included in Figs \ref{fig:spectra}) and \ref{fig:2D_spectra}, while those of the remaining galaxies can be found in the Appendix \ref{sec:app} (Figs \ref{fig:other_spectra} and \ref{fig:2D_spectra_rest}). 
These galaxies are likely quiescent though we cannot exclude the presence of obscured star formation. {Their D4000 breaks are broadly consistent with a $\sim 0.5$ Gyr-old population \citep{Gobat2012}, which favours the galaxies to be quiescent rather than lulling.} 
The two passive galaxies ($\rm P_N$ and $\rm P_S$ in Table \ref{tab:sample}) are located North/South of J1652 at projected distances of $\sim 40$ arcsec, with the Southern source being very close in redshift ($z=2.946$) to the quasar residing at $z=2.948$, while the Northern source is significantly offset ($z=2.931$). 

Further, we identify {two AGN candidates with $\log \oiii/\Hbeta > 0.8$ based on the grizli fits} described in Section \ref{sec:redshifts}. These are the two most Eastern sources in the field (IDs 5/$\rm LR_{1}$ and 6/$\rm LR_{2}$ in Table \ref{tab:sample}, corresponding to the 4th and 5th rows in Fig \ref{fig:other_spectra}). Their elevated $\oiii/\Hbeta$ ratio {likely places them in the AGN branch} of the \citet{BPT1981} (BPT) diagram at $z=3$ \citep{Kewley2013} even in the absence of \Halpha\ and \sii\ measurements, which are not available in our observations. The most Eastern source has a ratio of $\log \oiii/\Hbeta = 0.98$. In the case of negligible dust attenuation, it would have ${\rm O32} = \log \oiii/ \oii = 0.77$ and ${\rm R23} = \log \ (\OIII+\OII)/\Hbeta = 1.04$, consistent with $z=3-4$ Ly$\alpha$ emitters but exceeding the typical $z=3-4$ Lyman break galaxies \citep{Nakajima2016}. These values are further consistent with those in $4<z<10$ galaxies from the GLASS and CEERS spectroscopic surveys (Fig 1 in \citealt{Calabro2024}), indicating a strongly ionised and low-metallicity interstellar medium (ISM). Assuming a typical extinction $A_V = 0.7$ mag at redshift $3$ \citep{Cullen2018} slightly changes the line ratios to ${\rm R23}=1.05$ and ${\rm O32}=0.6$. The other AGN candidate instead has line ratios consistent with the Lyman break galaxies in \citet{Nakajima2016}: ${\rm O32} = 0.34$ ($0.2$ for $A_V = 0.7$ mag), $\log \oiii/\Hbeta = 0.81$ and ${\rm R23} = 0.98$ ($1.01$ for $A_V = 0.7$ mag). {In Table \ref{tab:fluxes}, we report both the observed line fluxes as well as those corrected for dust using our fiducial assumption of $A_V = 0.7$ mag.}

Our observations also yield two sources with broad \oiii\ lines blended with \Hbeta\ (IDs 10 and 12 in Table \ref{tab:sample}; rows 6 and 8 in Fig \ref{fig:other_spectra} cont.). We obtain \oiii\ FWHM of $1170, \ 1240$ km/s, respectively, which could be due to intrinsically broad lines or spatially extended emission. 
The cutout images show extended, potentially disturbed systems, and we cannot rule out mergers (though we note that the galaxy South-West from ID 10 appears to be an interloper in the foreground at $z=2.4$). Alternatively, line broadening could be caused by  rotation or extended gas flows.

\subsection{Structure of the protocluster}
\label{sec:PC_structure}

\begin{figure*}
    \sidecaption
    \includegraphics[width=11.7cm, trim={0.1cm 2.7cm 0.5cm 1cm}, clip]{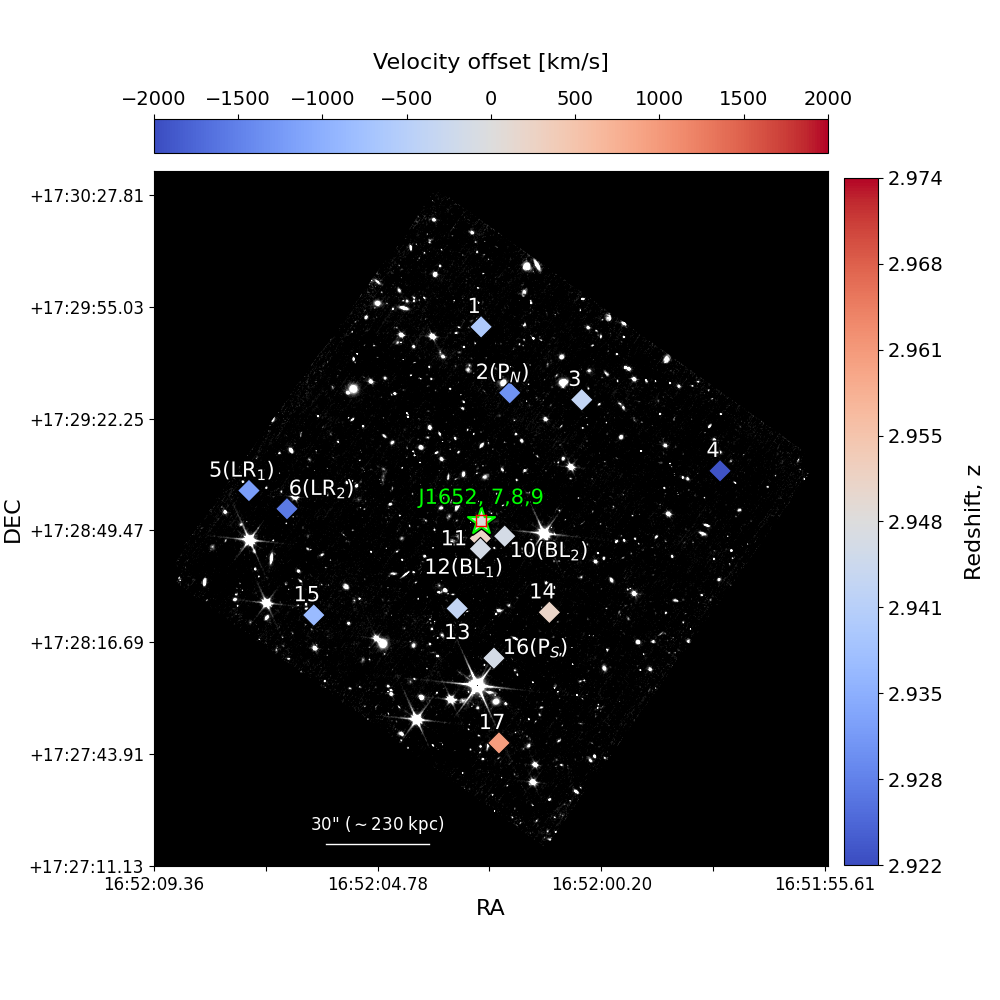}
    \caption{Overview of the protocluster field targeted by JWST/NIRISS. The spectroscopically confirmed cluster members (diamond symbols) are colour-coded by the (projected) velocity offset from the central source as well as redshift. The $3''$x$3''$ field of view of the JWST/NIRSpec observations - which contains the quasar and $3$ galaxies within a projected distance of $<20$ kpc - is overplotted in red in the centre.\\}
    \label{fig:protocluster_field}
\end{figure*}

We present a 2D view of the protocluster structure in Fig \ref{fig:protocluster_field}. The central star symbol covers both the quasar J1652 and the three additional galaxies from \citet{Wylezalek2022b} within the $3''$x$3''$ NIRSpec IFU field of view (red square). The member galaxies are marked by diamond symbols superimposed on the direct image, and their colour-coding reflects the redshift or corresponding velocity offset from the quasar, which roughly spans [-2000, +1000] km/s. The most blueshifted sources fall along the East-West direction, while redshifted targets are located southwards from the quasar. There is {an increased density} around J1652, with $6$ galaxies {(IDs 7,8,9,10,11,12)} falling within a projected distance of $70$ kpc from the quasar. Given this high galaxy density and the large mass of J1652 ($\log M_\star \sim 10.9$; see \citealt{Chen2026}), the quasar should reside approximately at the centre of the structure. {Thus, our observations uncovered a protocluster core. We compare the latter to the Distant Red Core at $z \sim 4$ containing 10-11 dusty star-forming galaxies within $280 \times 300$ kpc$^2$ \citep{Oteo2018} and SPT2349-56 at $z = 4.3$, which comprises 14 galaxies within $130 \times 130$ kpc$^2$ (\citealt{Miller2018}).
Within a region of $320 \times 230$ kpc$^2$, we find 10 galaxies (including IDs 13, 14 and 16), such that the density is comparable to that of the Distant Red Core, but lower than SPT2349-56. } 

\begin{figure}
    \centering
    \includegraphics[width=0.85\linewidth]{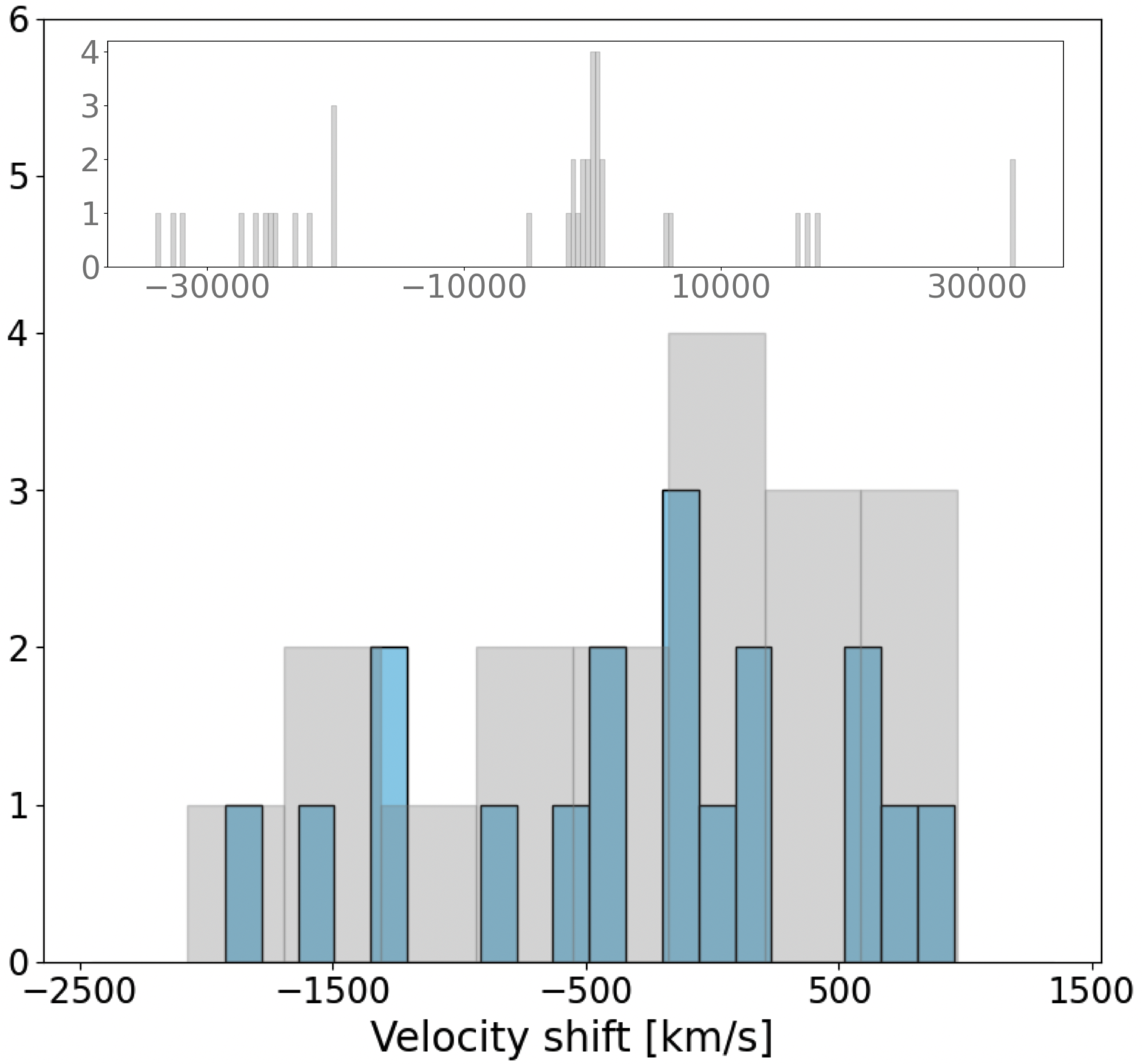}
    \caption{Distribution of the line-of-sight velocity offset with respect to the central quasar. {The main plot is limited to the protocluster members, while the inset plot includes the full redshift range $2.5 < z < 3.5$ over which our observations cover \oii , \oiii , and \Hbeta . The grey histograms illustrate} the impact of our adopted redshift uncertainty of $0.005$, which corresponds to $380$ km/s, the width of the grey bins. All member galaxies fall within a range of [$-2000$, $+1000$] km/s with respect to the quasar. There is a peak around the quasar, but the distribution is skewed with an excess of blueshifted sources compared to redshifted ones, which may result from the structure not being virialised yet. The velocity distribution together with the spatial distribution in Fig \ref{fig:protocluster_field_3D} suggest that the protocluster may be undergoing a merger of two haloes. }
    \label{fig:hist_velshift}
\end{figure}

We present a distribution of the line-of-sight {(LOS)} velocity offsets from the quasar in Fig \ref{fig:hist_velshift}, using $20$ bins of $\sim 140 {\rm km/s}$ each. The conservatively adopted redshift uncertainty of $dz = 0.005$ corresponds to a velocity shift of $380$ km/s at $z = 3$. To illustrate the impact of this uncertainty, we include a coarser grey histogram binned to $380$ km/s. {We further include an inset plot using the broader redshift range of $2.5<z<3.4$, where our observations cover the \oii, \oiii\ and \Hbeta\ lines, showing that the protocluster clearly stands out from the field via its density. The LOS velocity distribution of the protocluster members} is uneven, with almost double as many blueshifted sources relative to the quasar as redshifted ones. There is a clear peak around J1652, while the most blueshifted galaxies could potentially be falling into the structure. 
Although the distribution appears skewed by eye, suggesting that the structure is not virialised yet, a simple Kolmogorov–Smirnov test or Shapiro–Wilk test do not reject the null hypothesis that the velocities are sampled from a Gaussian given the low number statistics. We therefore proceed to combine the spatial and spectral information, which may provide deeper insights.

\begin{figure*}
    \centering
    \includegraphics[width=0.75\textwidth, trim={0.35cm 0cm 0.35cm 0cm}, clip]{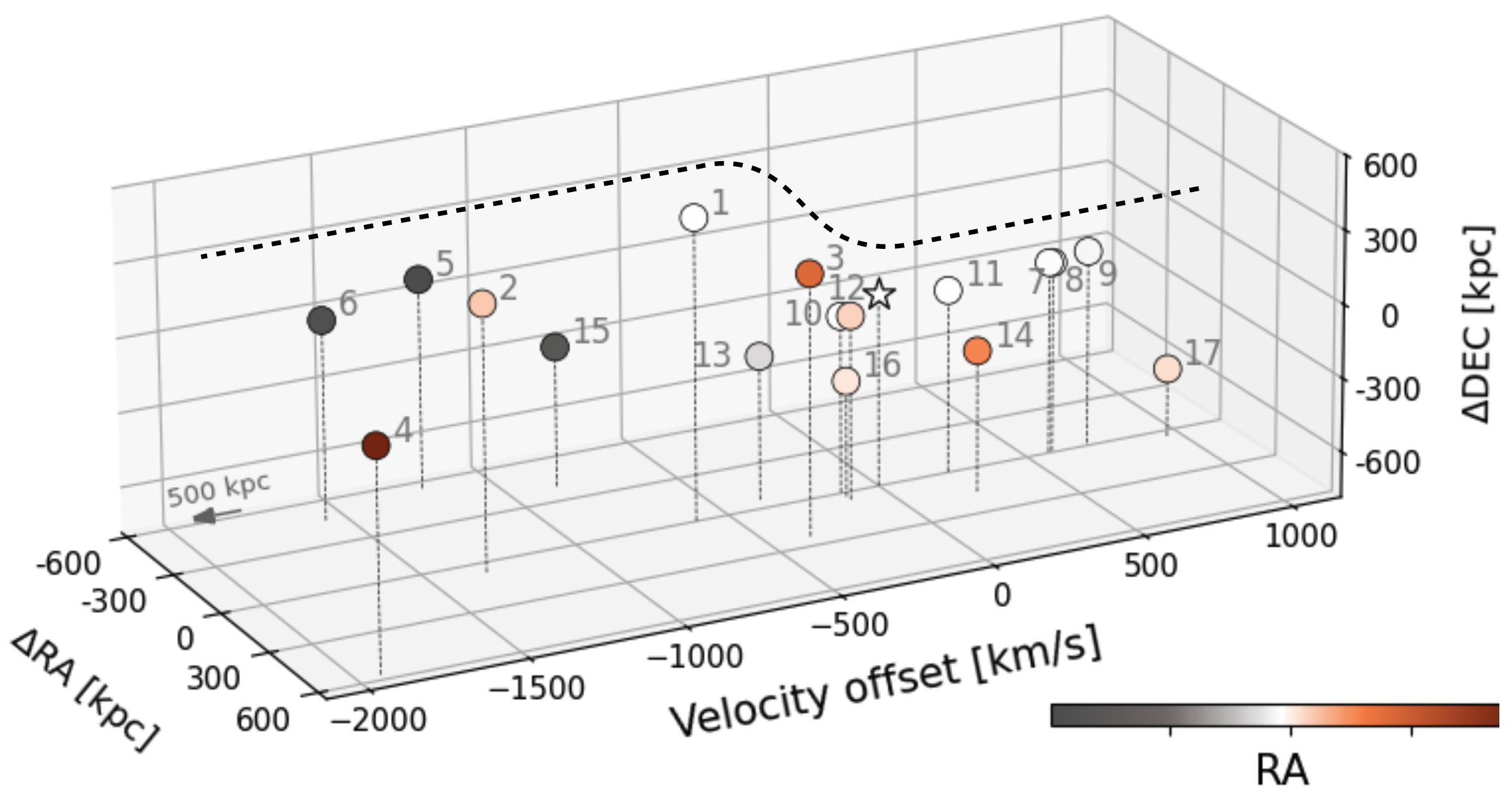}
    \caption{3D visualisation of the protocluster's structure. The quasar is marked by a star symbol, and {the IDs from Table \ref{tab:sample} are annotated}. {The colour scale reflects the RA (centred on the quasar) as a visual aid to assess the distribution}, and the axes denote the spatial offset w.r.t to the quasar in $\rm kpc$ along the North-South ($\rm \Delta DEC$) and East-West directions ($\rm \Delta RA$), as well as the LOS velocity shift. A LOS distance of 500 proper kpc is indicated by a grey arrow (if interpreting the observed redshifts entirely as spatial separation). The system may consist of two merging substructures, one in the North (at large negative velocity offsets) and one in the South (at smaller or positive velocity offsets), leading to the step-like structure sketched by the dashed line. 
    }
    \label{fig:protocluster_field_3D}
\end{figure*}

In Fig \ref{fig:protocluster_field_3D}, we visualise the 3D geometry (where the axis of velocity offset could also be partially interpreted as distance along the line of sight). {The colour-coding reflect the RA only as a visual aid}. The $2$ passive galaxies are annotated as $\rm P_N$ and $\rm P_S$. The densest region is located around the quasar, which is highlighted by a star symbol. Strikingly, the Northern sources are all significantly blueshifted, reaching almost $-2000$ km/s, while the Southern galaxies have predominantly positive or small negative velocity offsets with respect to the quasar, and are generally more clustered. Motivated by the step-like structure along the velocity axis, we dub the system ``The Step protocluster''. The anisotropic structure clearly indicates that the protocluster has not yet reached a dynamically relaxed state. 
As \citet{Wylezalek2022b} noted, the large velocity span between the quasar and its three closest companions, of the order $\sim 700$ km/s, is moreover difficult to explain with a single halo, and may point instead to a merger of two haloes. 
As we will discuss in Section \ref{sec:Comparison}, the 3D distribution supports the halo merger picture. Such a scenario could also potentially enhance the galaxy merger activity, contributing to the formation of the extended stellar tidal tail seen in the NIRSpec data and HST continuum.

\subsection{Member galaxy properties and overdensity}
\label{sec:Galaxy_properties}

\begin{figure}
    \centering
    \includegraphics[width=\linewidth, trim={0.4cm 0cm 0.3cm 0cm}, clip]{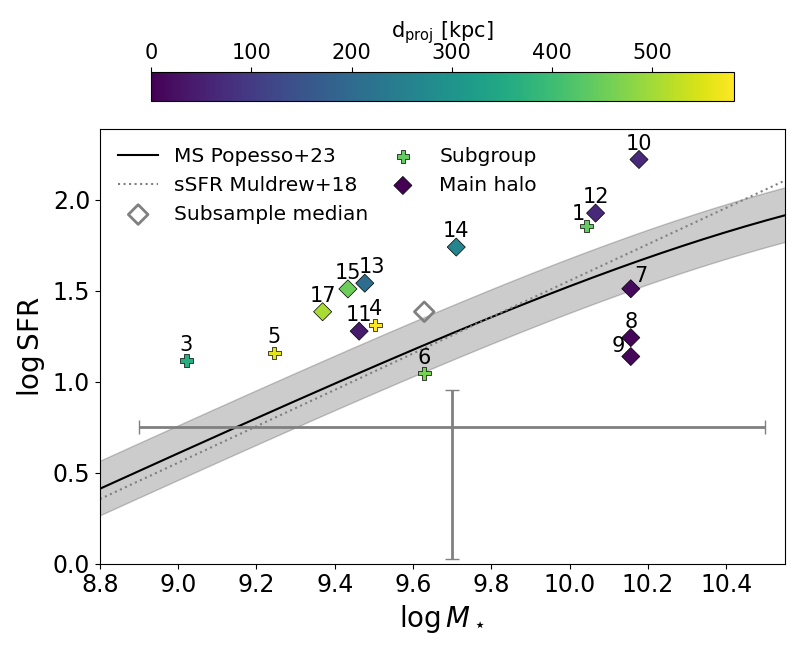}
    \caption{{SFR-$M_\star$ plane with the $z = 3$ star-forming Main Sequence \citep{Popesso2023}. The grey open diamond shows the median SFR and $M_\star$ of our sample (excluding the passive galaxies and quasar). We use our best estimates of the individual SFRs and stellar masses to tentatively add the individual member galaxies, colour-coded by their projected distance from the quasar. The grey cross represents the large uncertainties due to the assumed mass-to-light ratio, dust extinction and metallicity. Diamond symbols denote galaxies visually assigned to the main halo in Fig \ref{fig:protocluster_field_3D}, and cross symbols those in the Northern subgroup. The dotted line represents the specific star formation rate predicted by \citet{Muldrew2018} (for $h=0.7$). The least massive galaxies appear to belong to the potentially infalling subgroup, while sources closer to the quasar seem to be more massive on average. The star formation is generally slightly elevated by a median offset of $0.3$ dex compared to the Main Sequence.} }
    \label{fig:MS_plot}
\end{figure}

Our sample of 18 protocluster members spans a redshift spread of 0.038 (with redshift uncertainties of $dz=0.005$; see Section \ref{sec:redshifts}), corresponding to $9.5$ proper Mpc if interpreted in terms of spatial separation, leading to a maximum volume of $9.8$ Mpc$^3$. 
We thus derive a lower limit on the galaxy number density $\phi$ of $1.84$ Mpc$^{-3}$, which we compare to that of the field by integrating the stellar mass function at z=3 from \citet{Weaver2023} down to $\log \ M_\star / M_\odot = 9.14$, the approximate mass of the least massive galaxy in our sample, as we discuss below. This exercise yields $\phi_{\rm field} = 0.125 \ {\rm Mpc^{-3}}$ using the proper volume. 
The galaxy overdensity $\delta$ thus corresponds to a factor $\gtrsim 15$ {for the full field of view (1 Mpc$^2$), $>100$ for the ``core region" with 10 galaxies within $320 \times 230$ kpc$^2$ similar to the Distant Red Core \citep{Oteo2018}, and $>1600 / 8000$ for the $7/4$ galaxies (including the quasar) within $70^2 / 20^2$ kpc$^2$}. {Based on complementary VLT/MUSE data (PI: Assef), Gatto et al. (in prep.) also confirm an overdensity of a factor $>10$ in Ly$\alpha$ emitters.} 
A comparison to literature is complicated by the fact that the area sampled can vary significantly across different studies ($\sim 1 - 5^2$ proper Mpc$^2$). With that limitation in mind, only a handful of systems with $\delta > 10$ have been identified within the redshift range $2 < z < 4$ \citep{Cucciati2014, Lemaux2014, Dey2016, Polletta2021, Pensabene2024, Shimakawa2024}, making the Step a particularly rare system.

To assess the evolutionary stage of the structure, we compare the star-forming properties of the protocluster members to the field. We roughly estimate the median SFR and stellar mass within the protocluster. For the stellar masses, we calculate the restframe B-band luminosity using the fits described in Section \ref{sec:redshifts} (including the lines), and we assume a mass-to-light ratio of $\log (M_\star / L) = -0.3$ in solar units for a 1 Gyr-old stellar population (\citealt{Maraston2005}), as well as a typical extinction $A_V = 0.7$ mag at redshift $3$ \citep{Cullen2018}. 
Using the \citet{Calzetti2000} dust law, we thus obtain a crude median stellar mass estimate of $\log \langle M_\star / M_\odot \rangle \sim 9.7$. However, the $M_\star / L$ ratio could plausibly vary between $0.16$ and $1.6$ given its dependence on age and metallicity (\citealt{Maraston2005}), leading to an uncertainty of $\pm 0.5$ dex, and assuming $A_V = 0 - 1.5$ mag adds an additional $\pm 0.3$ dex in uncertainty. For the $3$ galaxies within $<20$ projected kpc of the quasar uncovered by NIRSpec data, whose NIRISS spectra are strongly contaminated by the quasar, we use the lower mass limit of $\log M_\star = 9.8$ following \citet{Wylezalek2022b}. The two passive galaxies have approximate stellar masses of $\log \ M_\star / M_\odot \sim 10.2, \ 10.4$. That makes them the second and fourth most massive galaxy in the protocluster, if taking our mass estimates at face value (with the most massive galaxy being the quasar at $\log  M_\star / M_\odot = 10.9$ from the decomposition of \citealt{Chen2026}).  {While it is not possible to draw strong conclusions based on two sources only, we nevertheless derive an estimate of the passive fraction and tentatively compare it to the field.} Based on UV-to-IR SED modelling of galaxies from the COSMOS-Web survey \citep{Casey2023}, \citet{ArangoToro2025} find that $\sim 5$\% of $z = 3$ galaxies with $\log M_\star  / M_\odot > 9.5$ are passive in the field (their Fig 6). Based on our stellar mass constraints, we identify $13$ protocluster members with $\log M_\star  / M_\odot > 9.5$, including the two passive sources, implying a quiescent fraction {of $15$\%,} about three times higher than in the field, {though we note again the low number statistics}. {Taken at face value, this result matches the predicted quenched fraction from the semi-analytical model in \citet{Muldrew2018} (for $\log  M_\star / M_\odot = 9.5$ galaxies in the main halo) and is consistent with the Magneticum cosmological simulations, which predict a passive fraction of $\sim 20 \pm 15 \%$ in $z=3$ protoclusters with halo masses $\log M_{\rm halo}/M_\odot = 13-13.5$ (\citealt{Remus2023}; see Section \ref{sec:halo_mass} for a discussion of the halo mass). Yet,} observations of protoclusters {with a quiescent population} at similar redshifts have only recently begun to emerge, and only a handful of these systems are known to date (\citealt{Kubo2021}, \citealt{Shi2021}, \citealt{McConachie2022}, \citealt{Ito2023}, \citealt{Naufal2024}, \citealt{TanakaM2024}, {\citealt{Kakimoto2026}}). The protocluster environment has likely contributed to the formation of the passive galaxies in the Step system, as well as to the stellar mass build-up of the quasar which has a host galaxy mass of $\log M_\star/M_\odot \sim 10.9$ \citep{Chen2026, Bertemes2025}. Recent photometric JWST surveys of the GOODS-N and GOODS-S fields indeed showed that the brightest and reddest galaxies preferentially reside in overdense environments \citep{Helton2024}. {At $z=5.7$, \citet{Morishita2025} recently presented two overdensities with an enhanced fraction of galaxies with weak emission lines and evolved stellar populations, while \citet{Witten2026} find preferentially redder UV slopes and stronger Balmer breaks in the member galaxies of a $z=7.9$ protocluster.} This is consistent with predictions from simulations in which the first quenched galaxies evolve in dense environments where they experience more AGN feedback due to earlier and more rapid black hole growth \citep{KurinchiVendhan2024}.

We proceed to estimate SFRs from the \citet{Kennicutt2012} relation. The \citet{Kennicutt2012} conversion is based on Solar metallicity, whereas stellar metallicities at $z=3$ are typically $\sim 0.8$ dex lower than those of nearby galaxies \citep{Sommariva2012}, but with a larger scatter. As a result, the SFR may in reality be a factor $\sim 1.8$ lower than the estimate based on Solar metallicity (see Table 1 in \citealt{Bicker2005}) due to the difference in the number of ionising photons. On the other hand, galaxies in protoclusters are known to evolve faster and are thus expected to be more metal-rich than the field. Based on recent JWST observations, \citet{Wang2021} reported (super) Solar metallicities in the CGM of a massive radio galaxy at $z=4.5$. {From a theoretical point of view, protoclusters in the cosmological simulations suite Magneticum are typically enriched to solar metallicities at $z \sim 4$ \citep{Remus2023}.} We therefore continue by adopting the \citet{Kennicutt2012} conversion without attempting to correct for metallicity which cannot be easily derived from our observations. We convert the observed \Hbeta\ flux into \Halpha\ assuming an intrinsic ratio of $2.85$ (case B recombination) for star-forming galaxies. For AGN candidates, we derive an upper limit to the SFR using $\Hbeta /\Halpha = 3.1$ for the narrow-line region \citep{Halpern1983}.   
In the case of negligible dust, we obtain a median SFR of $11 \ M_\odot/{\rm yr}$, but with our fiducial assumption of $A_V = 0.7$ mag and a median stellar-to-nebular attenuation ratio of $0.67$ \citep{Shen2025}, the recovered median SFR is $22 \ M_\odot/{\rm yr}$\footnote{As the NIRISS spectra of the $3$ closest quasar companions are completely contaminated by the quasar, we instead used the NIRSpec IFU data to fit their \Hbeta\ line following the setup described in \citet{Bertemes2025}. For the quasar itself, we assume an SFR of $200 M_\odot / {\rm yr}$ as derived by \citet{Vayner2023} in non-AGN spaxels, which may be an underestimate of the total SFR in the quasar.}. 

{In Fig \ref{fig:MS_plot}, we show the recovered positions of star-forming protocluster member galaxies in the SFR-$M_\star$ plane, with the median of the star-forming subsample shown as a grey open diamond. The figure does not show the central quasar, which is a massive ($\sim 10^{11} M_\odot$; \citealt{Chen2026, Bertemes2025}) starburst forming $\gtrsim 200 M_\odot/{\rm yr}$ (\citealt{Vayner2023}). Overall, the protocluster is placed} somewhat above the $z=3$ star-forming Main Sequence from \citet{Popesso2023} with a median offset $\rm \langle \log \ SFR - \log \ SFR_{MS}(M_\star) \rangle = 0.3$ (excluding the passive galaxies), though the uncertainties are dominated by our assumption of the mass-to-light ratio, which may plausibly vary by $\pm 0.5$ dex and sway the stellar masses by the same amount.  If using the  Main Sequence from \citet{Merida2026} instead (based on a non-parametric star formation history shape), the median offset increases to $\rm \langle \log \ SFR - \log \ SFR_{MS}(M_\star) \rangle = 0.6$. {We use diamond symbols to mark those galaxies that we visually assign to the main halo around the quasar (Fig \ref{fig:protocluster_field_3D}). Cross symbols mark galaxies belonging to the potentially infalling subgroup, which appears to contain the two least massive galaxies of the structure. The colour-coding reflects the projected distance to the quasar, suggesting that sources closer to the quasar are more massive. On a side note, we highlight a challenge in the comparison with simulations. For instance, the semi-analytical models of \citet{Muldrew2018} (grey dotted line) and the Magneticum cosmological simulations (\citet{Remus2023} Fig 7) predict protocluster galaxies to lie on the Main Sequence. The latter work suggests that this is plausibly due to the {instantaneous} SFR being underpredicted, while averaging over longer timescales leads to realistic SFRs (and as a consequence, stellar and gas masses).}

With two strong AGN candidates in addition to the quasar, the AGN fraction in the protocluster is at least $11$\% ($19$\% if not counting the two passive galaxies since their lack of optical lines would prevent us from detecting any potential AGN activity). The incidence and mass function of AGN selected by emission line ratios remain poorly constrained at $z =2-3$. Therefore, in order to compare our measured AGN fraction in the protocluster to the field, we instead use the X-ray AGN fraction in the field. To this end, we integrate the $z\sim2$ X-Ray AGN host galaxy mass function from \citet{Bongiorno2016} and the mass function of star-forming galaxies from \citet{Weaver2023} down to $\log \ M_\star / M_\odot = 9.14 $ (best mass estimate for the least massive galaxy in our sample), and divide the resulting number density of X-Ray AGN by that of star-forming galaxies. We derive an X-ray AGN fraction in the field at $z=2$ for comparable stellar masses of $\sim 5$\%. We caution that different AGN selection techniques may yield different AGN fractions. Also, our AGN candidates are based on an extremely strong \oiii/\Hbeta\ ratio only since our data does not cover \Halpha\ and \sii. With these limitations in mind, our results are consistent with an enhancement of a factor $2$ or more in the AGN fraction of the protocluster compared to the field. There have been several reports of elevated AGN fractions in protoclusters \citep{Lehmer2009, Digby-North2010, Gatica2024, Macuga2019}, although other works have not found any indications of a significant enhancement \citep{Macuga2019}. 

{Altogether, based on the characteristics discussed in this Section, our observations are consistent with the evolutionary scenario proposed by \citet{Shimakawa2018} (their Fig. 13). In this scheme, the initial rapid growth phase in which protoclusters are fed by cold accretion streams concludes around $z \sim 3$, and it is followed by a maturing phase ($2<z<3$) characterised by powerful AGN activity, pre-heating of the intracluster medium, core collapse, and the formation of a red sequence. Given its dense core, powerful central quasar and potentially elevated AGN fraction, and the presence of two passive galaxies, the Step structure is showing several of these maturing signatures.}

\subsection{Halo mass and normalised star formation} 
\label{sec:halo_mass}

We constrain the halo mass via different approaches. First, we estimate the dynamical mass via the velocity dispersion of the protocluster member galaxies. In virialised systems, the velocity distribution of the galaxies is set by the halo mass, but protoclusters are generally non-virialised. As a result, the recovered dynamical mass estimate should be regarded as an upper limit to the halo mass.  As an example, if non-virialisation increases the observed velocity dispersion $\sigma_{200}$ by a factor $1.5$ compared to a virialised state, the halo mass will be overestimated by $0.5$ dex given that it scales approximately with $\sigma_{200}^3$. 

Following \citet{Finn2005}, we estimate the virial radius to be $0.44$ Mpc from the observed 1D velocity dispersion of all cluster members (with the caveat that our observations could miss galaxies outside the NIRISS FoV). The LOS velocity dispersion of all member galaxies at smaller projected distances is $\sigma_{200} \sim 500$ km/s, which corresponds to a halo mass $M_{\rm halo}$ of $\sim 5 \times 10^{13} \Msun$ following \citet{Evrard2008, Sifon2013}\footnote{Perturbing the sources within our conservatively adopted uncertainty of $380$ km/s could technically increase this estimate by a factor $4$, but a halo mass exceeding $10^{14} M_\odot$ is unrealistic at this redshift.}. 
Given that the Step system potentially represents a merger of two haloes, as we will discuss in more detail in Section \ref{sec:PC_structure}, we repeat the exercise by separating the Northern from the Southern sources. 
The Northern sources cluster around the passive galaxy, which is the most massive among them, has an intermediate velocity and lies approximately at the centre of the substructure, as shown in Fig \ref{fig:protocluster_field_3D}. This leads to a total halo mass estimate of $7.2 \times 10^{13} \Msun$ ($5.7 \times 10^{13} \Msun$ for the Northern structure only due to its large spread in velocities). 
As a final constraint, we place our stellar mass estimates on the $z=3$ stellar mass - halo mass relation from \citet{Shuntov2022}, yielding a summed halo mass of $\log M_{\rm halo}/ M_\odot = 13.2$ (see e.g. \citealt{Laporte2022}) with our adopted $A_V=0.7$, or $\log M_{\rm halo}/ M_\odot = 13.1$ for $A_V=0$, which could potentially be underestimated given the limitations by the field of view and sensitivity of our observations. Other caveats include large uncertainties in the stellar masses, as well as the stellar mass - halo mass relation at high redshifts. 

All things considered, our recovered range of $13.1 < \log  M_{\rm halo}/ M_\odot < 13.8$ is consistent with the Step protocluster potentially being a progenitor to the most massive, Coma or Virgo-like clusters at $z \sim 0$ (for which \citet{Chiang2013} recover a progenitor mass range of $13.2 \lesssim \log  M_{\rm halo}/ M_\odot \lesssim 13.7$ at $z=3$). 
Literature reports of similarly large halo mass estimates at $z>2$ (e.g. \citealt{Cooke2008, Kuiper2011, Lemaux2014, Kubo2016, Oteo2018, Champagne2021, Ito2023, Darvish2024}) are not as rare as expected given the low number of such massive cluster systems observed at low redshift. This may be a result of both the observational uncertainties involved in determining halo masses of unvirialised systems, as well as the difficulty in predicting the theoretical evolution of massive structures due to large variations in halo growth histories \citep{Chiang2013}, especially with increasing redshifts. Recent results from the Magneticum cosmological simulation suite suggest that by $z=4.3$, the most massive protocluster do not necessarily evolve into the most massive clusters by $z=0$ \citep{Remus2023}. {In the latter work, several protoclusters' tracks (in their Fig. 10) exhibit halo masses at $z=3$ which are consistent with our estimated range. By $z=0$, they evolve to have diverse halo masses of typically several $10^{14}$ to $\gtrsim 10^{15} M_\odot$. }

\begin{figure*}
    \sidecaption
    \includegraphics[width=11.5cm, trim={0.4cm 0.3cm 0.4cm 0cm}, clip]{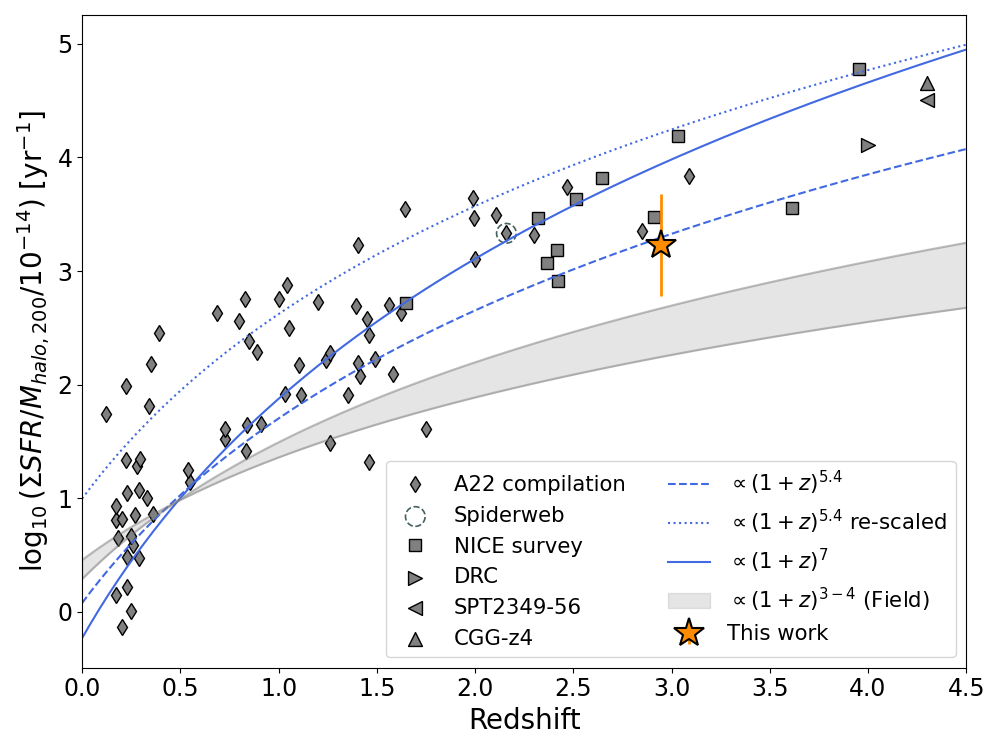}
    \caption{{Redshift evolution of the halo-normalised total SFR, with the Step protocluster superimposed (star symbol). The grey symbols denote (proto-)clusters from the literature with IR-based SFR measurements. The diamonds mark the compilation presented in \citet{Alberts2022}, {and we list the corresponding individual references in the main text}. {We add a dashed circle to highlight the $z=2.16$ Spiderweb protocluster included in this compilation.} Targets from the NOEMA formIng Cluster survEy (NICE) are shown as squares \citep{Sillassen2024, Zhou2025}. The triangles show the Distant Red Core (DRC; \citealt{Long2020}) at $z=4$, SPT2349-56 at $z = 4.3$ \citep{Hill2022} and CGG-z4 at $z = 4.3$ \citep{Brinch2025}. The blue lines show different redshift evolutions derived for (proto-)clusters: $\propto (1+z)^{7}$ (solid; \citealt{Cowie2004, Geach2006, Biviano2011}) and $\propto (1+z)^{7}$ (dashed/dotted line for $M_{\rm halo} \sim 14.5 / 14$; \citealt{Popesso2015b}). The redshift evolution of the field, $\propto (1+z)^{3-4}$, is indicated by the grey shaded region. Based on our \Hbeta-based estimate of the total SFR, which may miss obscured star formation, the Step appears to lie on the lower envelope of the distribution, suggesting a typical or slightly suppressed level of star formation. } }
    \label{fig:Alberts_plot}
\end{figure*}

Having tentatively constrained the halo mass, we can further characterise the star-forming state of the protocluster by estimating the halo mass-normalized total SFR, $\Sigma {\rm SFR} / M_{\rm halo}$, which traces the efficiency of the stellar mass build-up. As mentioned in the introduction, the bulk of known high-z protoclusters are more active sites of star formation than the field. \citet{Alberts2022} conducted a compilation of IR (proto-)cluster studies from the literature out to $z \sim 3$, which supports that $\Sigma {\rm SFR} / M_{\rm halo}$ in these systems follows a steep redshift evolution $(1+z)^{5.4-7}$, compared to only $(1+z)^{3-4}$ in the field {(see also \citealt{Zhou2025})}. {In Fig \ref{fig:Alberts_plot}, {we reproduce Fig 7 from \citet{Alberts2022} to} put the Step protocluster in the context of the literature, {using their compiled values of the halo-normalised SFR from the studies of \citet{Haines2013}, \citet{Santos2014, Santos2015}, \citet{Popesso2015a, Popesso2015b}, \citet{Ma2015}, \citet{Alberts2016}, \citet{Casey2016}, \citet{Strazzullo2018}, and \citet{Lacaille2019}. We add a dashed circle to mark the Spiderweb structure, which is included in \citet{Casey2016} (taken originally from \citealt{Kurk2000}, \citealt{Kuiper2011} and \citealt{Dannerbauer2014}), and which we will discuss as a potential analogue in Section \ref{sec:Comparison}.} 
We also add the targets from the NOEMA formIng Cluster survEy (NICE; \citealt{Sillassen2024, Zhou2025}), as well as the Distant Red Core at $z=4$ \citep{Oteo2018, Long2020}, SPT2349-56 at $z = 4.3$ \citep{Miller2018, Hill2022}, and CGG-z4 at $z=4.3$ \citep{Brinch2025}. The datapoints at higher redshift indicate that these systems are consistent with the predicted redshift evolutions $\propto (1+z)^{5.4-7}$ (blue lines). }
In the Step protocluster, our estimates of $\Sigma {\rm SFR} / M_{\rm halo}$ range between $0.6 \times 10^{-11} \ {\rm yr^{-1}}$ (no dust, upper halo mass estimate of $10^{13.8} M_\odot$) and $4.8 \times 10^{-11} \ {\rm yr^{-1}}$ (for $A_V = 0.7$ mag, $\log M_{\rm halo} = 13.1$). However, we caution that {our \Hbeta - based SFRs could miss obscured star formation,} the SFR of the quasar may be severely underestimated, and higher amounts of dust than our fiducial assumption of $A_V = 0.7$ mag would further increase the total SFR. In terms of star-forming efficiency, despite the presence of the two quiescent galaxies, the Step structure {thus appears to be consistent with other $z=3$ protoclusters, or perhaps slightly less active}.

\subsection{A single halo, merger or two separate groups?: Comparison to Step analogues}
\label{sec:Comparison}

{As noted in the previous section, the Step protocluster is broadly consistent with other high-z protoclusters in terms of the star-formation efficiency (or alternatively mildly suppressed). However, it} stands out from typical systems at $z \sim 3$ {via the combination of its} large mass, broad velocity distribution, large central velocity span, presence of quiescent galaxies, and substructure in 3D position–velocity space. The latter raises the question of whether the Step is a connected system at all or instead a chance alignment of two non-interacting structures that do not collapse into a single descendant cluster by $z=0$. In this context, we compare the system to two rare protoclusters with similar properties{ - SSA22 at $z=3.09$ \citep{Steidel1998} and SpiderWeb at $z=2.2$ \citep{Pentericci2000} - and we consider additional high-z structures. }

The SSA22 system \citep{Steidel1998} at z $\sim 3$ is a massive overdensity with visible substructure consisting of two spatially separated blue- and red-shifted groups \citep{Topping2016}. The distribution of velocities in the system follows a broad and clearly double-peaked shape. SSA22 also hosts a massive quiescent galaxy \citep{Kubo2021}. Based on identifying similar systems in the Small MultiDark Planck (SMDPL) dark matter simulation \citep{Behroozi2013}, \citet{Topping2018} concluded that the SSA22 field likely represents a chance alignment of two non-interacting structures that do not collapse into a single descendant cluster by $z=0$. In the Step system, the number statistics are too low to confirm or reject a double-peaked distribution in the velocity histogram, but assuming that the Northern (IDs 1-6 and 15) and Southern structures represent distinct groups, they are separated by $\sim 1200$ km/s. In contrast, the red- and blueshifted groups in SSA22 are separated by $\sim 2200$ km/s in their peaks, significantly larger than the separation in the Step, while each having similar or slightly narrower velocity dispersions ($400-500$ km/s). 
{Moreover, perhaps one of the most distinctive characteristics of the Step is the presence of 4 galaxies within a projected search radius of $<12$ kpc in its red subgroup, which span a velocity range of $700$km/s between them. These kinematics are difficult to explain without a source of disturbance. In contrast, within SSA22's red or blue subgroup, there are no 2 galaxies within $<30$ kpc with a velocity difference exceeding $500$ km/s (using the public catalogue from \citealt{Mawatari2023}). If expanding the search radius to 70 kpc, the Step's densest region contains 7 galaxies spanning a velocity range of $\sim 850$km/s approximately continuously (with best-fit velocity shifts of [-130, -130, 0, 230, 580, 590, 710] km/s). For the same area in SSA22, the largest velocity span does reach $\sim 650$km/s, but this is only between a single galaxy pair, without any evidence for a third galaxy.} Thus, while the SSA22 is best explicable as a chance alignment of two structures \citep{Topping2018}, it is not clear if the same conclusion is applicable to the Step system. However, we proceed to consider a different analogue.

The SpiderWeb structure \citep{Kurk2000} is one of the most well-known and extensively studied protoclusters beyond $z=2$. At its core, it hosts a powerful and extremely massive radio galaxy (PKS 1138-262; \citealt{Seymour2007, Hatch2009}) surrounded by a large Ly$\alpha$ nebula. The extended environment is overdense in Ly$\alpha$ emitters \citep{Kurk2000}, H$\alpha$ emitters \citep{Kurk2004}, X-Ray sources \citep{Croft2005}, and NIR-selected quiescent galaxy candidates \citep{Kodama2007}. Deep HST observations reveal a web-like structure consisting of dozens of merging satellite galaxy candidates \citet{Miley2006}. {Based on zoom-in simulations, \citet{Saro2009} suggest that the SpiderWeb protocluster is likely a progenitor to a rich $z=0$ cluster with a halo mass in excess of $10^{15} M_\odot$.} Recent slitless spectroscopic follow-up observations {further} confirmed the existence of a red sequence with $11$ robust quiescent galaxies \citep{Naufal2024}. {The latter may be explained by the core of the structure exhibiting both a paucity of molecular gas and a lower efficiency of star formation, based on comparable $z=2.2$ haloes in the \textsc{DIANOGA} cosmological simulations \citep{Esposito2025}.}

Intriguingly, the kinematic properties of the SpiderWeb structure show several striking similarities to the Step. The analysis of \citet{Kuiper2011} revealed a very broad velocity distribution ranging from approximately -2000 to 1500 km/s, similar to the velocity span of the Step (Fig \ref{fig:hist_velshift}), with a total velocity dispersion of $\sim1000$ km/s compared to $\sim800$ km/s in the Step. There is a clear substructure, with blueshifted and redshifted regions being spatially separated {within a $\sim 130$ kpc aperture}. Further, the densest region of SpiderWeb contains 4 galaxies in a search radius of 30 kpc (including PKS 1138-262), reminiscent of the Step, though with an even larger velocity span of $>1000$ km/s between them (see Fig 2 in \citet{Kuiper2011}). The velocities follow a double-peaked shape with a separation of $1600$ km/s between the peaks of the two groups, somewhat closer to the $\sim 1200$ km/s spanned between the Step's Northern and Southern structures than the aforementioned $2200$ km/s in the SSA22 system. \citet{Kuiper2011} also conducted a comparison to the Millenium cosmological simulation \citep{Springel2005}, showing that a single massive halo is unlikely to reproduce the observed kinematics. A merger of two massive dark matter haloes in the simulation is able to temporarily increase the observed velocity shifts and dispersion by a factor $\sim 2$-$3$ for one timesclice at the first pass. The resulting peak velocity difference between the two merging groups is $\sim 1500$ km/s along the LOS, and the velocity dispersion reaches $\sim 700-900$ km/s, well matched to the SpiderWeb observations. The similarities in these properties between SpiderWeb and the Step system suggest that the latter is likewise a good candidate for a merger between two dark matter haloes. {Recent zoom-in simulations suggest that halo mergers may play a key role in protocluster evolution by promoting the early formation of the hot intracluster medium \citep{Bennett2026}. Indeed \citet{Tozzi2022b} recently confirmed diffuse thermal X-Ray emission in the core of SpiderWeb.}

{In our analysis, we have also considered other literature examples of high-z protoclusters as potential analogues to the Step (given the availability of spectroscopic redshifts for a significant number of members). Another case well-known for its prominent substructure is MRC 0316-257 at $z=3.1$ \citep{Venemans2005}. However, its study is complicated by the fact that the two subgroups are spatially offset at a projected distance of 1.4 Mpc, such that their observed LOS velocity difference of 1600 km/s is likely only a lower limit, and the observations may therefore be explained either by a merger or a non-interacting alignment \citep{Kuiper2012}. Similarly, the two $z=3.4$ protoclusters presented in \citet{McConachie2022} (MAGAZ3NE J095924$+$022537 and MAGAZ3NE J100028$+$023349) 
are separated by a projected distance of $\sim 8$ proper Mpc. 
The massive Hyperion system at $z=2.45$ is part of a much larger supercluster with 7 density peaks (Fig 3. in \citealt{Cucciati2018}) and is thus more complex than the Step. On the other hand, the Step protocluster's velocity distribution spanning $3000$ km/s exceeds those of SPT2349-56 at z=4.3 (1300 km/s; \citealt{Miller2018}) and the Distant Red Core at z=4 (2400 km/s; \citealt{Oteo2018}). While our comparison to literature is not exhaustive, it reinforces our conclusion that the Step is an exceptional system, and comparable analogues are accordingly sparse. 
}

{At lower redshift, there is} a system similar to the Step and SpiderWeb, namely around the luminous quasar PKS 0454-22 at $z\sim 0.5$. \citet{Helton2021} recently studied the environment around this source and noted that the kinematics and substructure could potentially be explained by a merger of two groups of galaxies. The recovered velocity distribution spans [$-1000$, $1500$] km/s, with a red and blue subgroup separated by $\sim 1200$ km/s. Within $25$ projected kpc of the quasar, there are two additional galaxies spanning $900$ km/s between them (six within $80$ kpc spanning $1100$ km/s), reminiscent of the environment in the Step and SpiderWeb. Galaxy groups that share similar kinematic properties and substructure are promising candidates for probing the merger-driven growth of massive haloes.

\section{Conclusions}

We have presented a newly discovered protocluster dubbed ``The Step'' at $z \sim 3$ unveiled by JWST/NIRISS grism observations targeting the overdense field around the luminous quasar and starburst J1652 ($z_{\rm spec} = 2.9489$). Previous NIRSpec IFU observations had already unveiled 3 galaxies within a projected radius of $<12$ kpc from the quasar at a velocity offset of $\sim 500-700$ km/s \citep{Wylezalek2022b}. {The NIRISS observations confirm this previously reported overdensity.} 
Our results are as follows:
\begin{itemize}
    \item We identify 18 robust member galaxies within [-2000, +1000] km/s of J1652, including 12 new sources based on \oii\ + \oiii\ emission and 
     2 massive quiescent galaxies with a clear D4000 break (consistent with a $\sim 0.5$ Gyr old stellar population) with no evidence for ongoing star formation (Fig \ref{fig:spectra}). Two members are AGN candidates based on optical line ratio diagnostics, and two other members show evidence for either intrinsically broad \oiii\ lines or extended emission. 
    \item The system is composed of two structures, which may be in the process of merging (discussed in Section \ref{sec:Comparison}), given the separation of Northern versus Southern sources in the 3D position-velocity space, creating a step-like pattern (Fig \ref{fig:protocluster_field_3D}). This may also explain the large velocity spread in the densest region ($700$ km/s between $4$ galaxies spanning projected distances of $<20$ kpc between each other), and enhanced galaxy merger activity may have caused the extended tidal tail detected in the stellar continuum \citep{Wylezalek2022b}.
    \item The protocluster seems to be in a transitional phase, given the coexistence of a starbursting quasar and two massive quiescent galaxies, implying a passive galaxy fraction $\sim 3$ times higher than in the field. Compared to the field, the star-forming galaxies still appear to show slightly elevated star formation for their masses, but the latter could vary by $\pm 0.5$ dex depending on the adopted mass-to-light ratio alone. 
    \item We derive a lower limit to the galaxy overdensity of a factor $\delta > 15$ within a $\sim 1$ Mpc$^2$ FoV, and estimate a large halo mass of $\log M_{\rm halo} / M_\odot = 13.1 - 13.8$. {With 10 galaxies within $320 \times 230$ kpc$^2$ ($\delta_{\rm core} > 100$), the core's density is similar to that of the Distance Red Core ($z=4$; \citealt{Oteo2018}).} The structure is thus consistent with being a progenitor of today's most massive, Coma-like clusters, though the evolution of individual haloes is challenging to predict \citep{Chiang2013, Remus2023}. For the given halo mass, the level of star formation is broadly consistent with other $z=3$ protoclusters, or slightly suppressed depending on assumptions (Section \ref{sec:halo_mass}).
    \item With two new AGN candidates (in addition to the quasar), we find tentative evidence for an enhancement in the AGN fraction ($10-20 \%$) relative to the field. However, the comparison is complicated by differences in AGN selection methods. 
\end{itemize}

{Our findings are consistent with several of the maturing signatures predicted from the evolutionary scenario proposed in \citet{Shimakawa2018} (Fig 13), in which the epoch of $z \sim 3$ marks the end of rapid protocluster growth fuelled by cold accretion streams. In this picture, the subsequent turnover is characterised by powerful AGN activity, pre-heating of the intracluster medium, core collapse, and the formation of red, ageing galaxies. The Step protocluster exhibits a dense core, a powerful central quasar (as well as a tentatively elevated AGN fraction), and two quiescent galaxies. It} constitutes one of few recently-discovered high-redshift protoclusters with {at least one quiescent galaxy} and/or signs of suppressed star formation (\citealt{Kubo2021}, \citealt{Shi2021}, \citealt{McConachie2022}, \citealt{Ito2023}, \citealt{Naufal2024}, \citealt{TanakaM2024}, {\citealt{Kakimoto2026}}) unveiled by sufficiently deep observations to detect line-less galaxies. {Altogether, these findings support a scenario where} high-z protoclusters are efficient sites for the formation of early massive and/or quiescent galaxies. This is consistent with photometric JWST surveys, which show that bright and red galaxies at high redshifts preferentially inhabit overdensities \citep{Helton2024}. The high AGN fraction in the Step system is broadly consistent with a picture where early passive galaxies experience accelerated growth and enhanced AGN activity in dense environments as suggested by cosmological simulations \citep{KurinchiVendhan2024}. 
All things considered, our findings contribute to the emerging picture that the redshift range $2 \lesssim z \lesssim 4 $ is a transitional regime for massive protoclusters {see also \citealt{Shimakawa2018})}, where these systems tip over from being the most active sites of star formation in the early Universe towards becoming the virialised clusters populated predominantly by massive, passive ellipticals that are observed at low redshift.

\begin{acknowledgements}
We thank the anonymous referee for helpful comments. CB and DW acknowledge support through an Emmy Noether Grant of the German Research Foundation, a stipend by the Daimler and Benz Foundation and a Verbundforschung grant by the German Space Agency (DLR). CB also acknowledges support from an additional DLR grant {through program 50OR2507}. This work is based in part on observations made with the NASA/ESA/CSA James Webb Space Telescope. The data were obtained from the Mikulski Archive for Space Telescopes at the Space Telescope Science Institute (STScI), which is operated by the Association of Universities for Research in Astronomy (AURA), Inc., under NASA contract NAS 5-03127 for JWST. These observations are associated with program \# GO-03807. Support for program \# GO-03807 was provided by NASA through a grant from STScI. REH acknowledges support by the German Aerospace Center (DLR) and the Federal Ministry for Economic Affairs and Energy (BMWi) through program 50OR2403 'RUBIES'. GN acknowledges support by the Canadian Space Agency under a contract with NRC Herzberg Astronomy and Astrophysics.
\end{acknowledgements}

\bibliographystyle{aa}
\bibliography{CB_bib_merged.bib}

\begin{appendix}

\section{Spectra of the cluster members and line fluxes}
\label{sec:app}

We present the spectra of the cluster members in Figs \ref{fig:spectra} and \ref{fig:other_spectra}. Further, we present the measured \oii, \oiii\ and \Hbeta\ line fluxes in Table \ref{tab:fluxes}.

\begin{figure*}[]
{
\ContinuedFloat
    \centering
    \includegraphics[width=0.62\linewidth, valign=m]{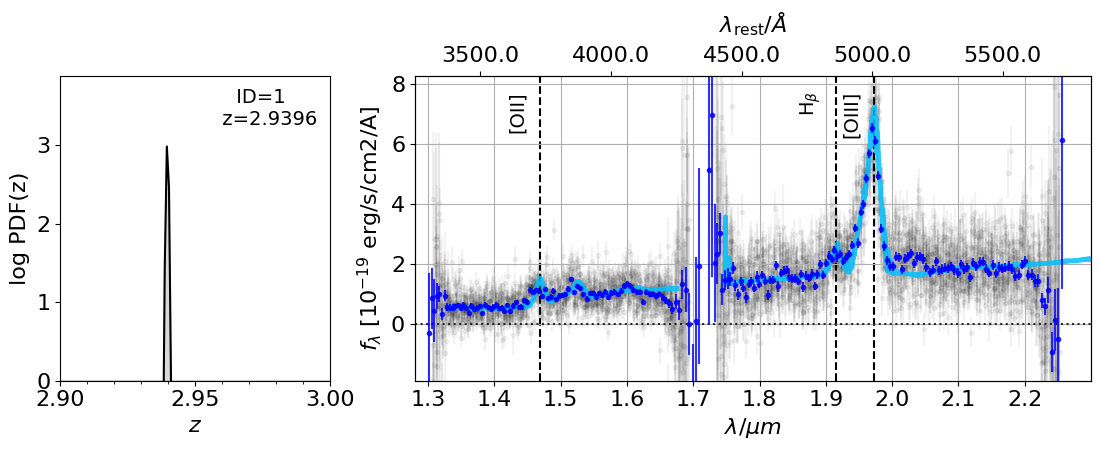}
    \includegraphics[width=0.23\linewidth, valign=m, trim={6cm 1.5cm 6cm 2cm}, clip]{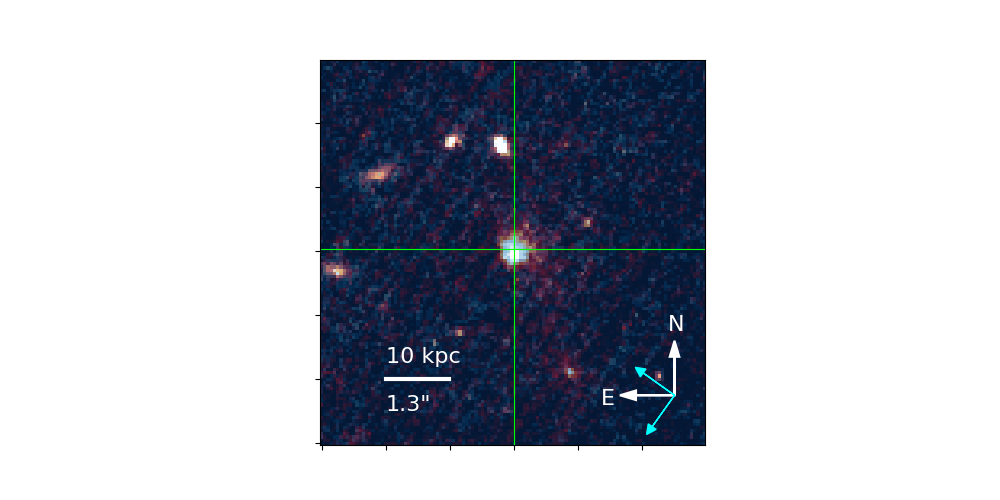}
    
    \includegraphics[width=0.62\linewidth, valign=m]{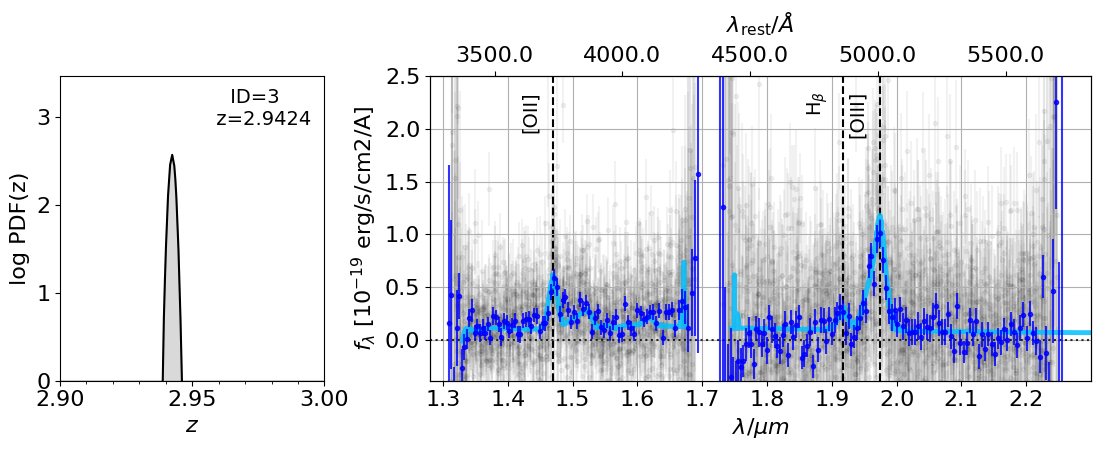}
    \includegraphics[width=0.23\linewidth, valign=m, trim={6cm 1.5cm 6cm 2cm}, clip]{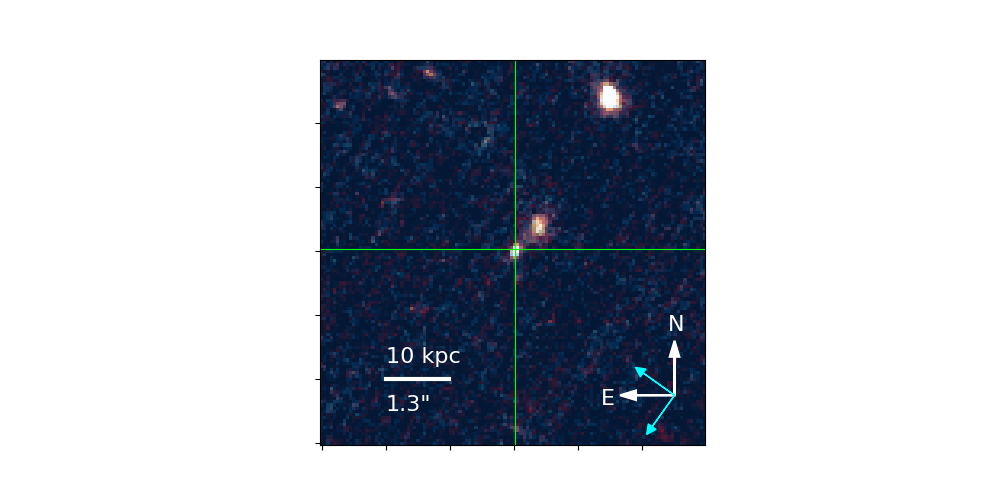}
    
    \includegraphics[width=0.62\linewidth, valign=m]{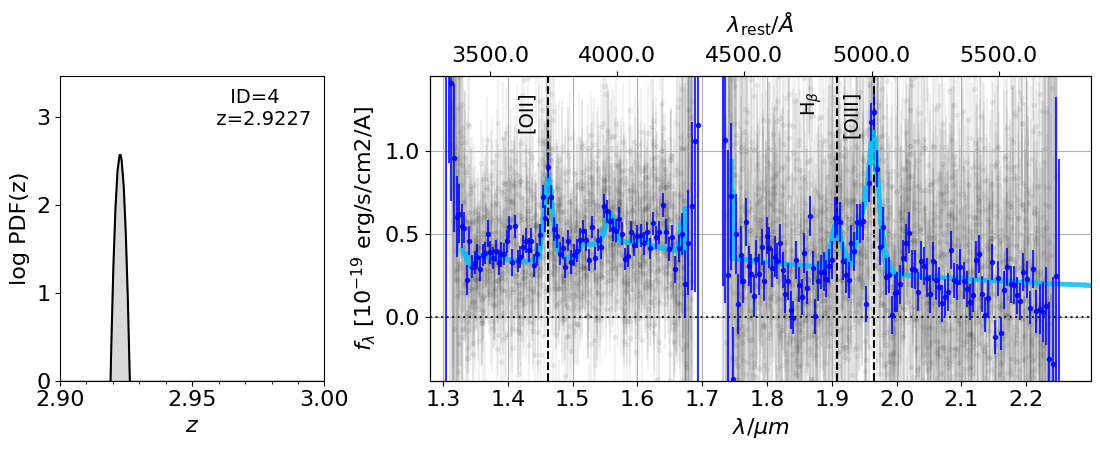}
    \includegraphics[width=0.23\linewidth, valign=m, trim={6cm 1.5cm 6cm 2cm}, clip]{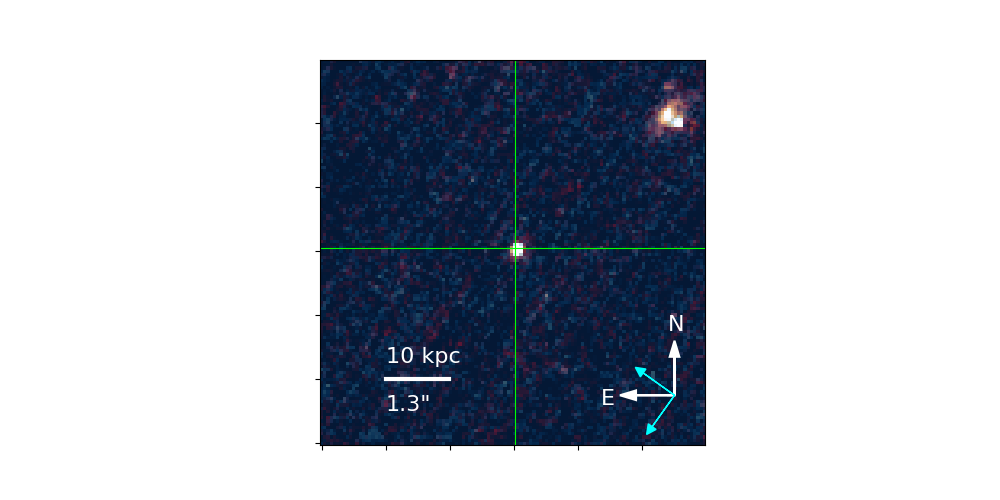}

    \includegraphics[width=0.62\linewidth, valign=m]{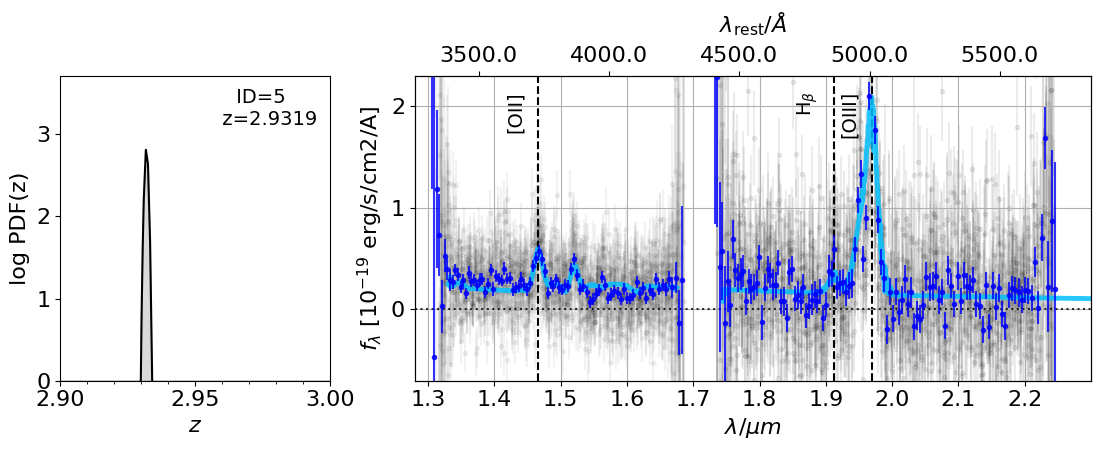}
    \includegraphics[width=0.23\linewidth, valign=m, trim={6cm 1.5cm 6cm 2cm}, clip]{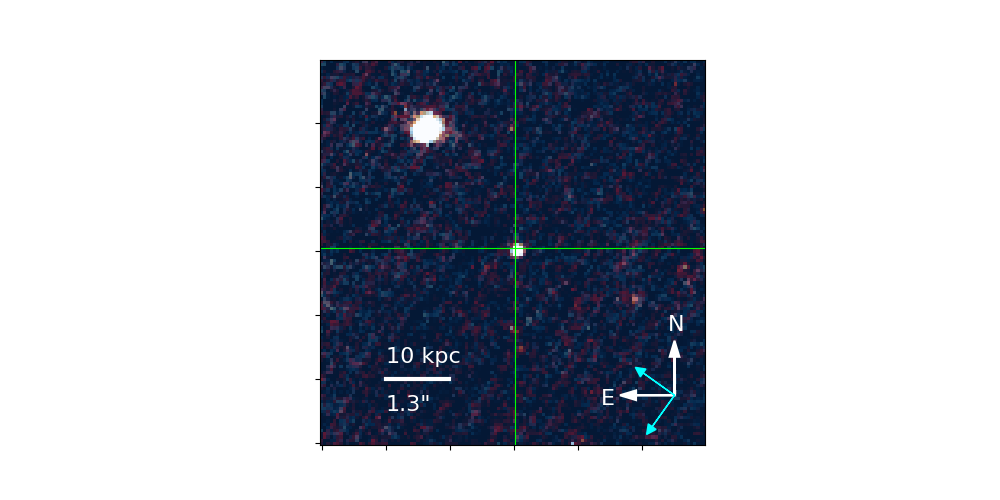}
    
    \includegraphics[width=0.62\linewidth, valign=m]{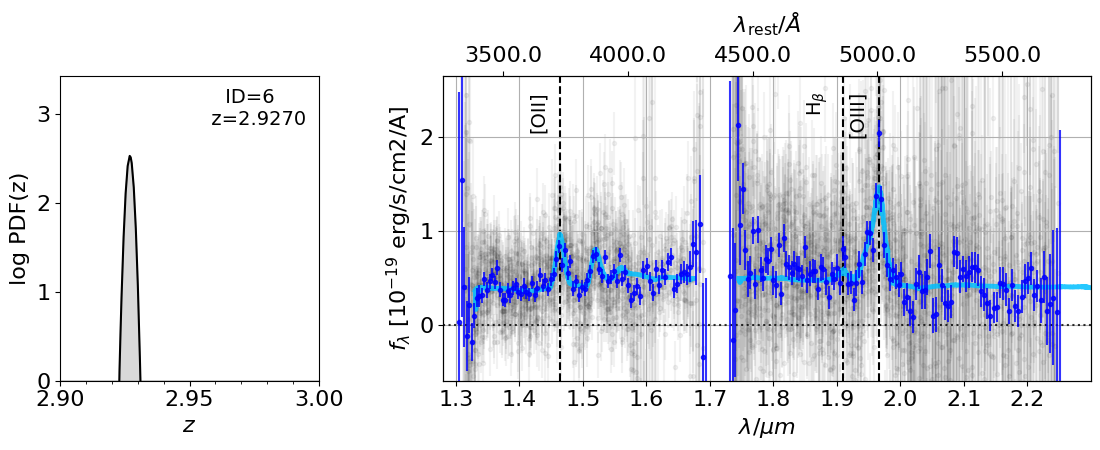}
    \includegraphics[width=0.23\linewidth, valign=m, trim={6cm 1.5cm 6cm 2cm}, clip]{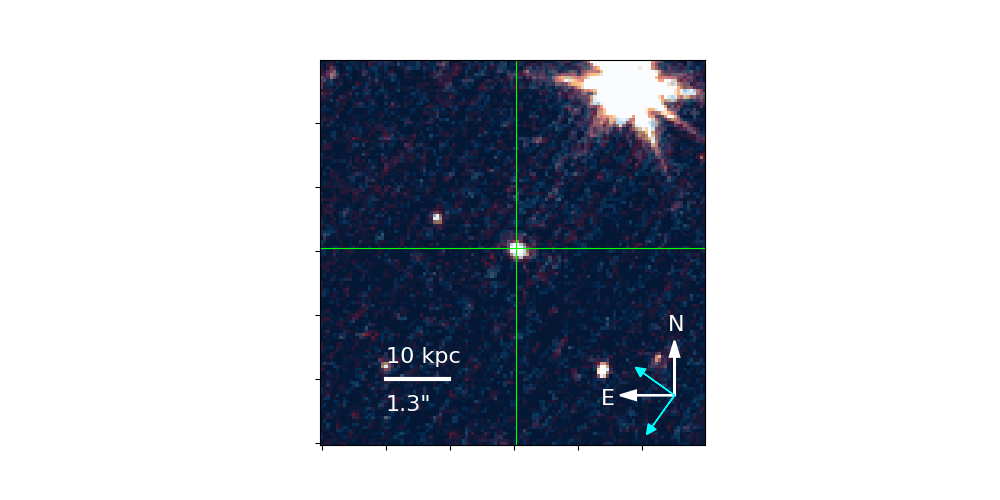}

\caption{NIRISS spectra of the protocluster member galaxies not shown in Fig \ref{fig:spectra}, excluding the three closest companions to the quasar uncovered by the JWST/NIRSpec IFU data due to contamination in the NIRISS data.}
\label{fig:other_spectra}
}
\end{figure*}

\begin{figure*}[]
{
\ContinuedFloat
    \centering
    \includegraphics[width=0.55\linewidth, valign=m]{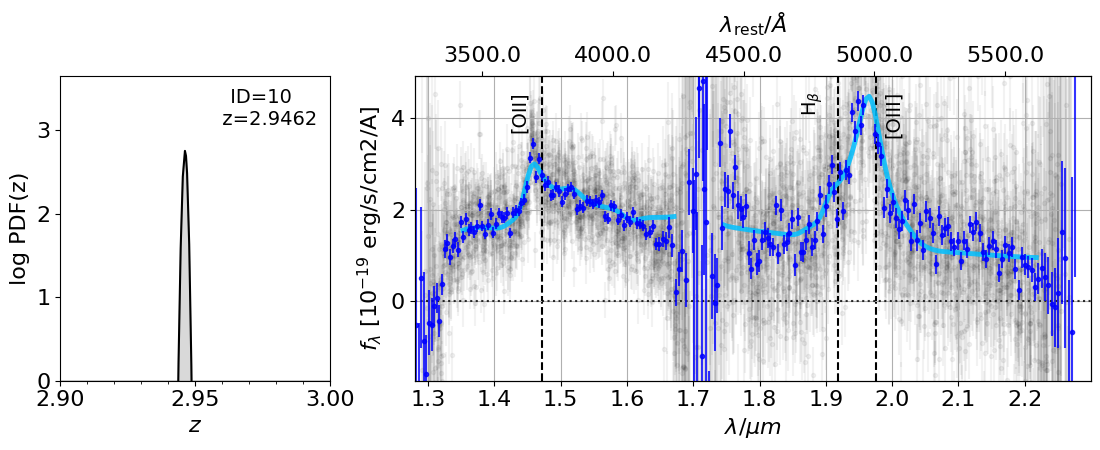}
    \includegraphics[width=0.23\linewidth, valign=m, trim={6cm 1.5cm 6cm 2cm}, clip]{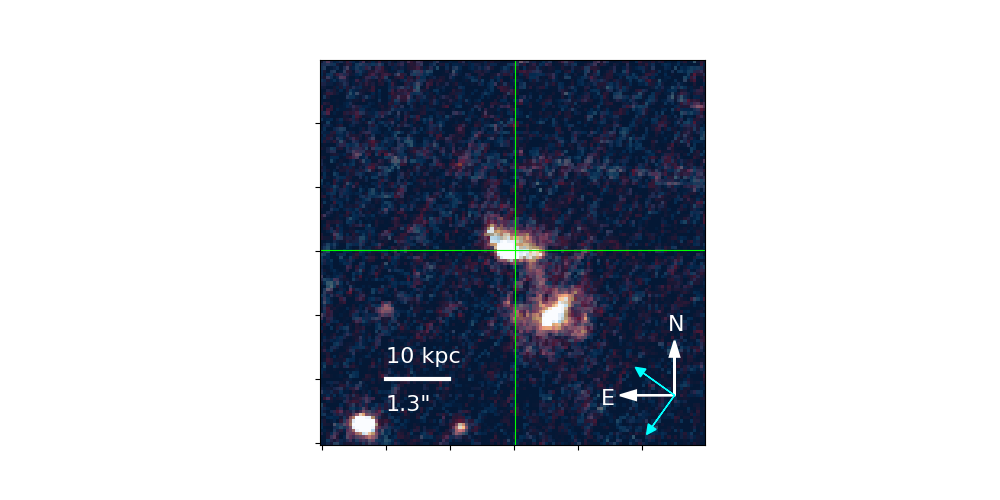}
    
    \includegraphics[width=0.55\linewidth, valign=m]{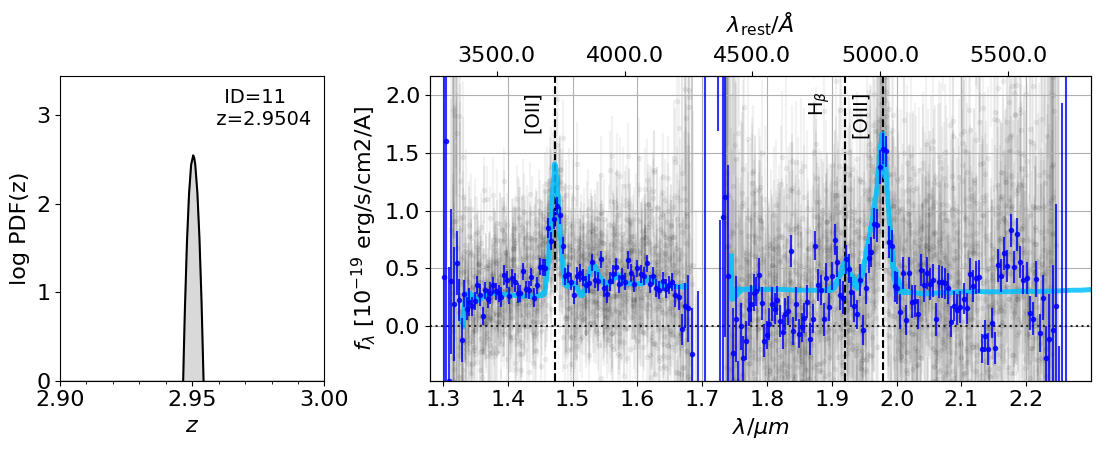}
    \includegraphics[width=0.23\linewidth, valign=m, trim={6cm 1.5cm 6cm 2cm}, clip]{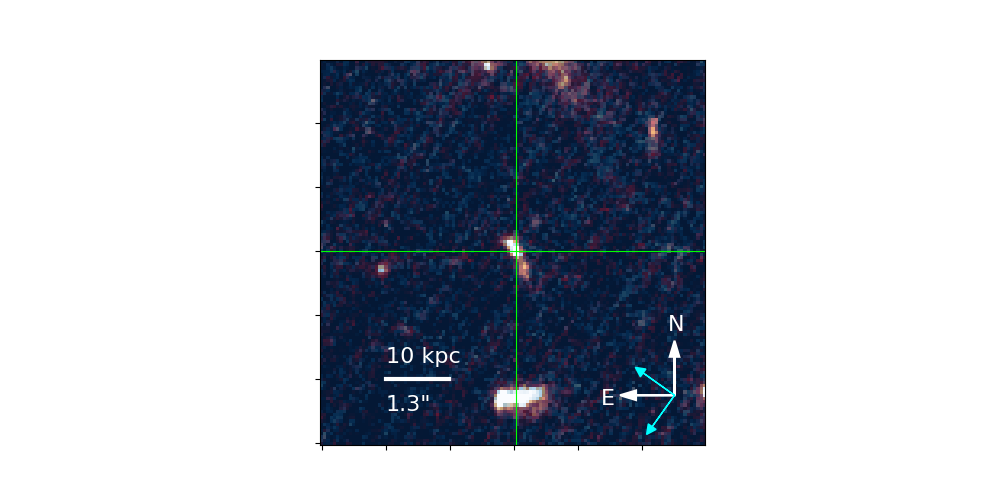}

    \includegraphics[width=0.55\linewidth, valign=m]{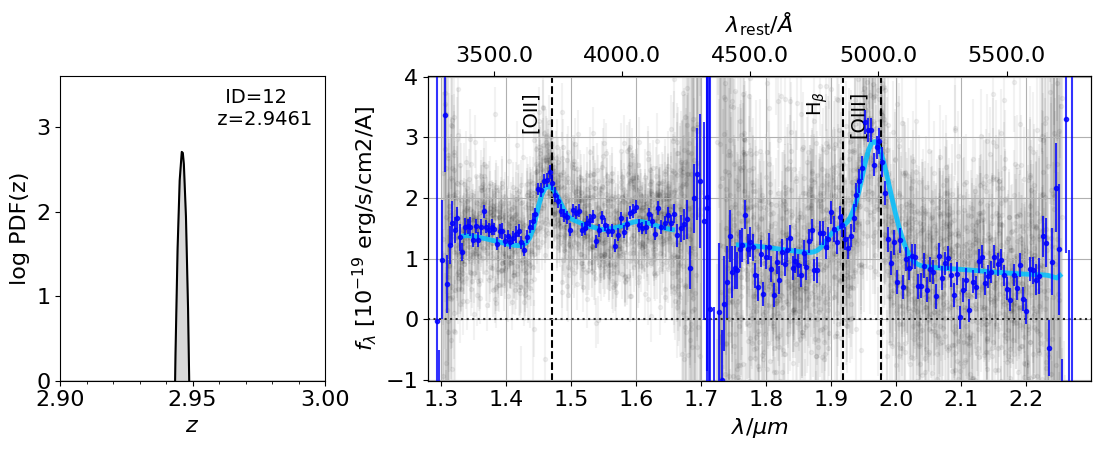}
    \includegraphics[width=0.23\linewidth, valign=m, trim={6cm 1.5cm 6cm 2cm}, clip]{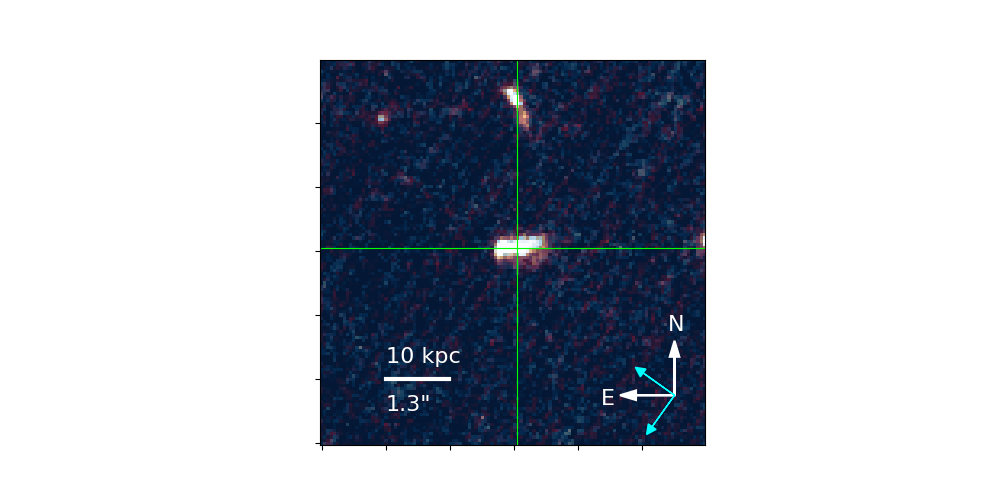}

    \includegraphics[width=0.55\linewidth, valign=m]{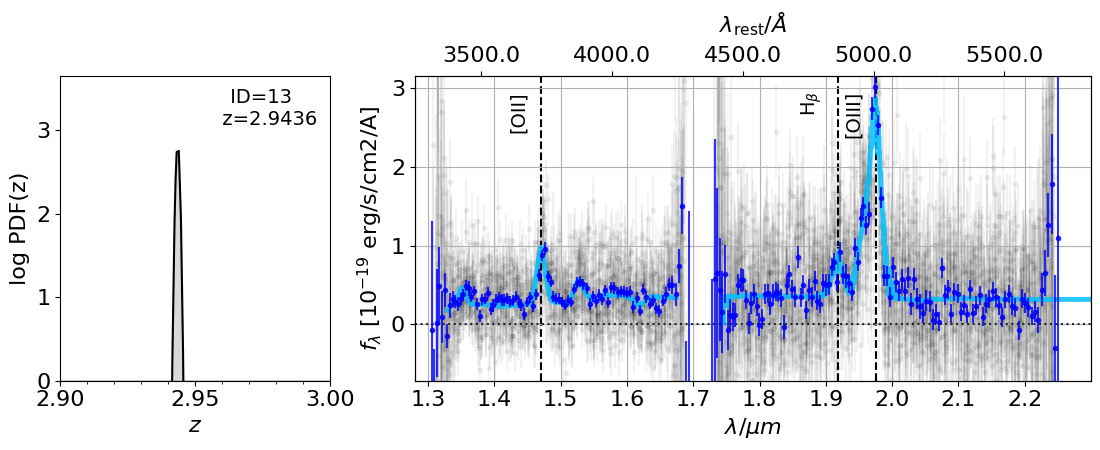}
    \includegraphics[width=0.23\linewidth, valign=m, trim={6cm 1.5cm 6cm 2cm}, clip]{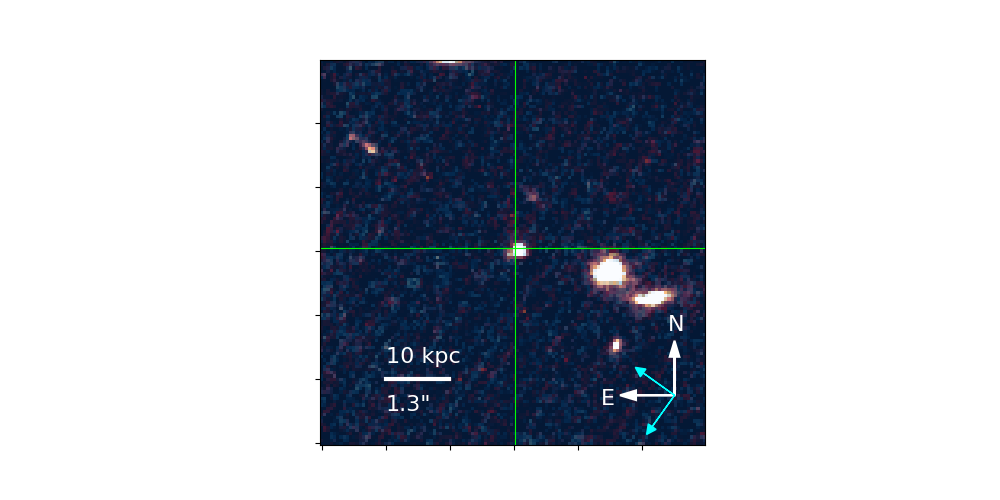}
    
    \includegraphics[width=0.55\linewidth, valign=m]{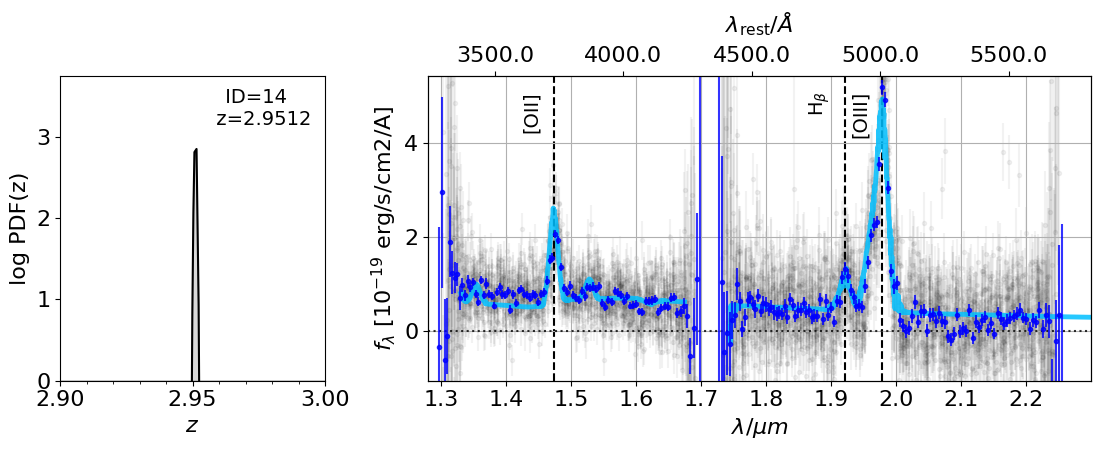}
    \includegraphics[width=0.23\linewidth, valign=m, trim={6cm 1.5cm 6cm 2cm}, clip]{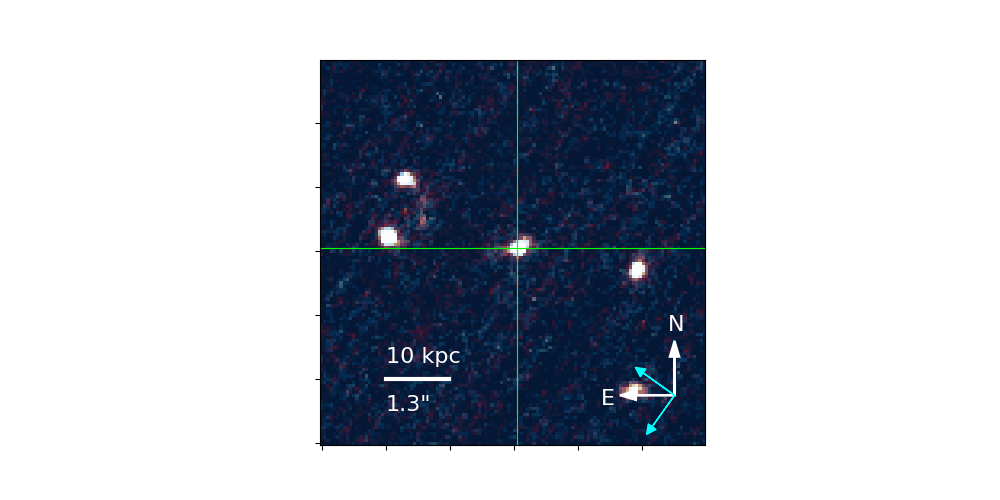}   
    \includegraphics[width=0.55\linewidth, valign=m]{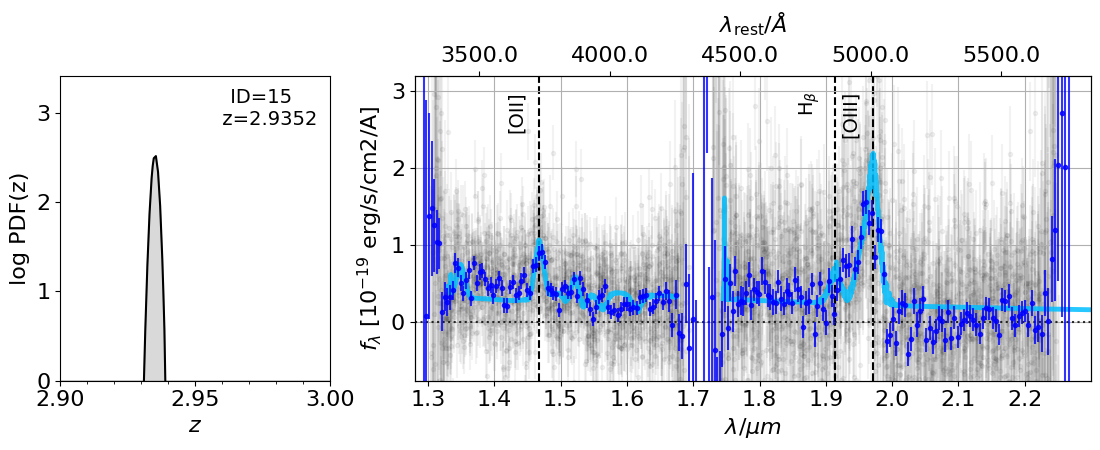}
    \includegraphics[width=0.23\linewidth, valign=m, trim={6cm 1.5cm 6cm 2cm}, clip]{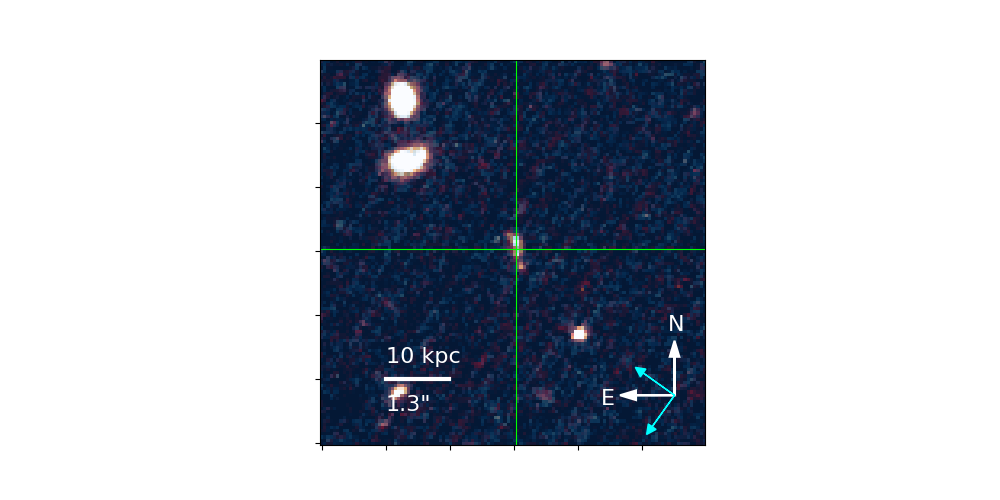}
\caption{continued.}
}
\end{figure*}

\begin{figure*}[]
{
    \begin{minipage}{\linewidth}
        \begin{minipage}{0.07\linewidth}
        \phantom{a}
        \end{minipage}
        \begin{minipage}{0.92\linewidth}
        \hspace*{1.5cm} F150W \hspace*{2.5cm} F200W  \hspace*{4cm} F150W \hspace*{2.5cm} F200W  \\
        \end{minipage}
    \end{minipage}
    \begin{minipage}{0.065\linewidth}
            Grism 1 \\[3.5ex]
            Grism 2 \\[3.5ex]
            Line      \vspace*{0.7cm}
    \end{minipage}
        \centering
    \begin{minipage}{0.93\linewidth}
        \includegraphics[width=0.47\linewidth, trim={0cm, 0.8cm, 0cm, 0cm}, clip]{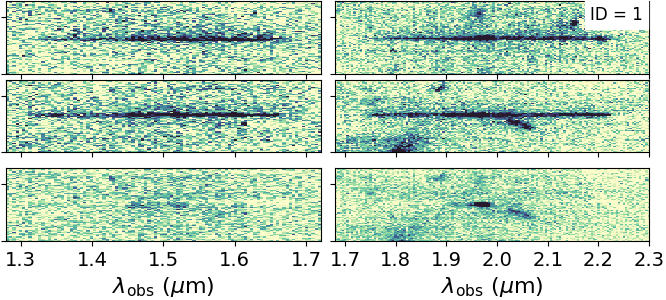}
        \vspace*{0.3cm}
        \begin{minipage}{0.05\linewidth}
        \phantom{a}
        \end{minipage}
        \includegraphics[width=0.47\linewidth, trim={0cm, 0.8cm, 0cm, 0cm}, clip]{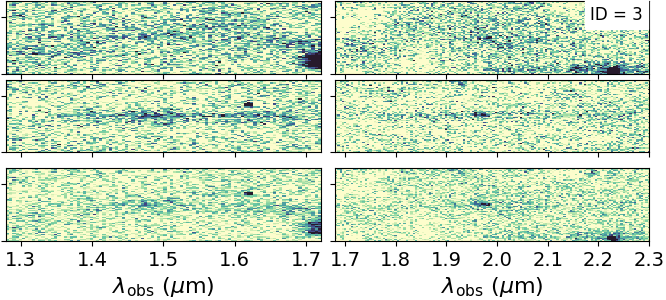}
    \end{minipage}
    \begin{minipage}{0.065\linewidth}
            Grism 1 \\[3.5ex]
            Grism 2 \\[3.5ex]
            Line      \vspace*{0.7cm}
    \end{minipage}
        \centering
    \begin{minipage}{0.93\linewidth}
        \includegraphics[width=0.47\linewidth, trim={0cm, 0.8cm, 0cm, 0cm}, clip]{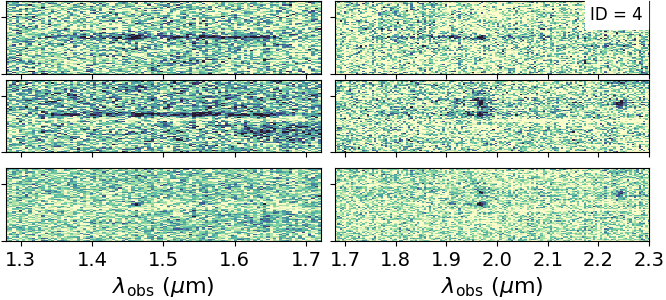}
        \vspace*{0.3cm}
        \begin{minipage}{0.05\linewidth}
        \phantom{a}
        \end{minipage}
        \includegraphics[width=0.47\linewidth, trim={0cm, 0.8cm, 0cm, 0cm}, clip]{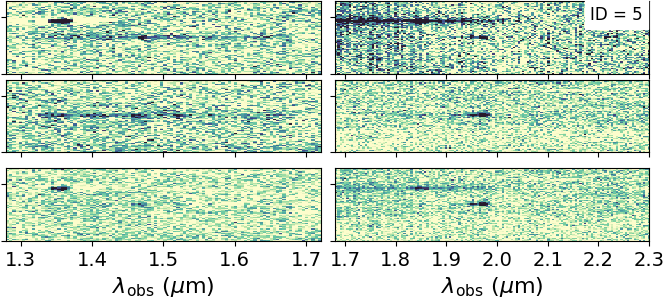}
    \end{minipage}
     \begin{minipage}{0.065\linewidth}
            Grism 1 \\[3.5ex]
            Grism 2 \\[3.5ex]
            Line      \vspace*{0.7cm}
    \end{minipage}
        \centering
    \begin{minipage}{0.93\linewidth}
        \includegraphics[width=0.47\linewidth, trim={0cm, 0.8cm, 0cm, 0cm}, clip]{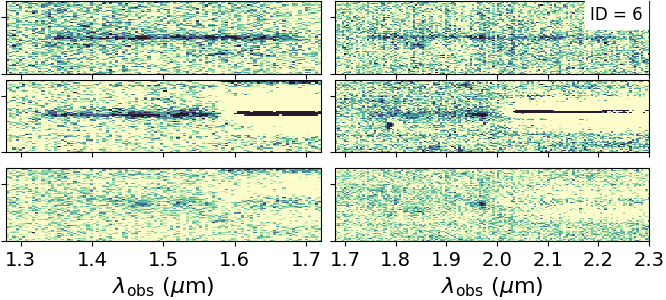}
        \vspace*{0.3cm}
        \begin{minipage}{0.05\linewidth}
        \phantom{a}
        \end{minipage}
        \includegraphics[width=0.47\linewidth, trim={0cm, 0.8cm, 0cm, 0cm}, clip]{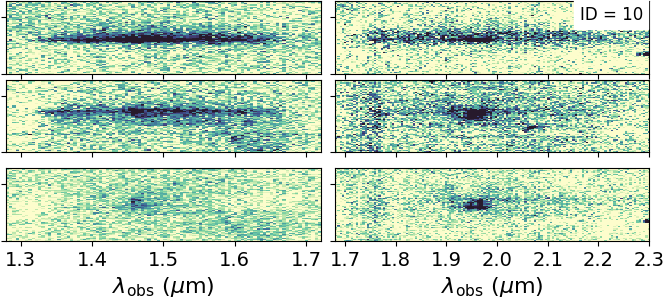}
    \end{minipage}

     \begin{minipage}{0.065\linewidth}
            Grism 1 \\[3.5ex]
            Grism 2 \\[3.5ex]
            Line      \vspace*{0.7cm}
    \end{minipage}
        \centering
    \begin{minipage}{0.93\linewidth}
        \includegraphics[width=0.47\linewidth, trim={0cm, 0.8cm, 0cm, 0cm}, clip]{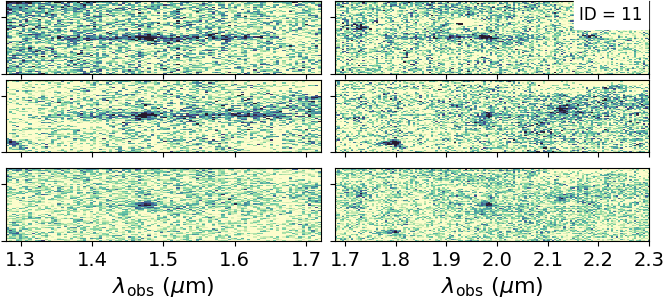}
        \vspace*{0.3cm}
        \begin{minipage}{0.05\linewidth}
        \phantom{a}
        \end{minipage}
        \includegraphics[width=0.47\linewidth, trim={0cm, 0.8cm, 0cm, 0cm}, clip]{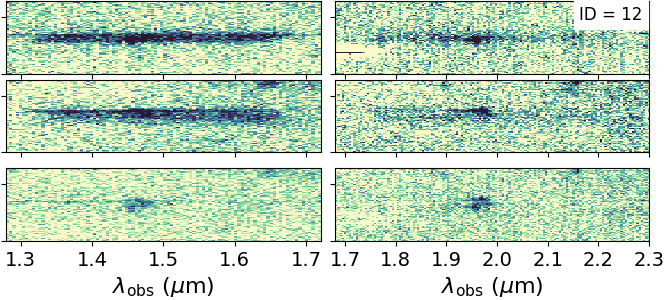}
    \end{minipage}

    \begin{minipage}{0.065\linewidth}
            Grism 1 \\[3.5ex]
            Grism 2 \\[3.5ex]
            Line      \vspace*{0.7cm}
    \end{minipage}
        \centering
    \begin{minipage}{0.93\linewidth}
        \includegraphics[width=0.47\linewidth, trim={0cm, 0.8cm, 0cm, 0cm}, clip]{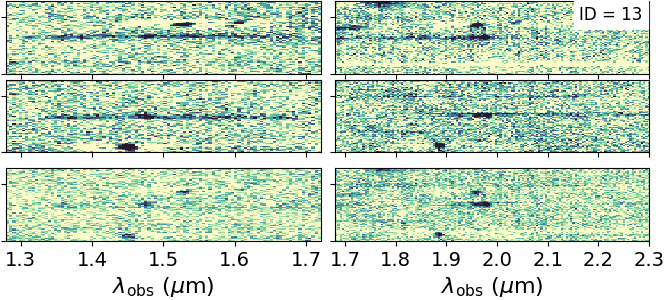}
        \vspace*{0.3cm}
        \begin{minipage}{0.05\linewidth}
        \phantom{a}
        \end{minipage}
        \includegraphics[width=0.47\linewidth, trim={0cm, 0.1cm, 0cm, 0cm}, clip]{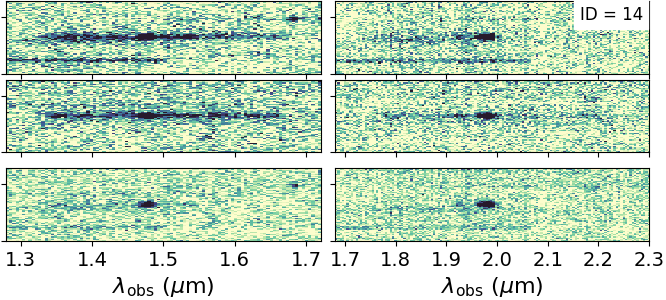}
    \end{minipage}
    \begin{minipage}{0.065\linewidth}
            Grism 1 \\[3.5ex]
            Grism 2 \\[3.5ex]
            Line      \vspace*{0.7cm}
    \end{minipage}
        \centering
    \begin{minipage}{0.93\linewidth}
        \includegraphics[width=0.47\linewidth]{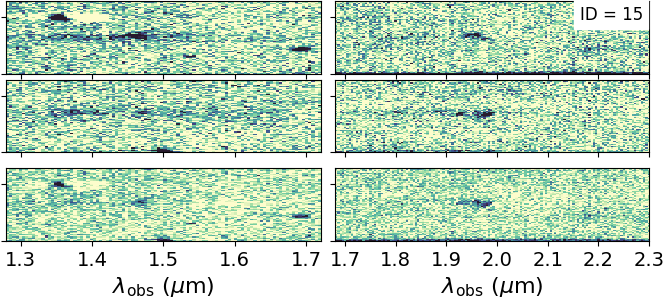}
        \begin{minipage}{0.52\linewidth}
        \phantom{a}
        \end{minipage}
    \end{minipage}   
\caption{2D NIRISS spectra of the {remaining} protocluster member galaxies {not shown in Fig \ref{fig:2D_spectra}}, excluding the quasar and its three closest companions uncovered by JWST/NIRSpec IFU data due to contamination in the NIRISS data. The left and right columns shows the spectra in the F150W and F200W filters, respectively, in observed frame. The first and second row show the spectra in both orthogonal grisms separately, while the third row shows the line emission after subtracting the modelled continuum based on a stack of the two grisms (without correcting for position angle). ID 6 (as annotated in the upper right corner) is affected by contamination from a nearby source in one grism, but only longwards of the lines. }
\label{fig:2D_spectra_rest}
}
\end{figure*}

\begin{table*}
\centering
\begin{minipage}{0.8\textwidth}
\caption{{\Hbeta\ , \oii\ and \OIII\ line fluxes for the protocluster member galaxies. }}
\label{tab:fluxes}
\begin{tabular}{ Sl Sl Sl Sl Sl Sl Sl}
\hline
ID & $\log \mathrm{L_{H\beta, \ obs}} $ & $\mathrm{\log L_{H\beta, \ corr}}$ & $\log \mathrm{L_{[OII], \ obs}}$ &  $\log \mathrm{L_{[OII], corr}}$ & $\log \mathrm{L_{[OIII]5007, \ obs}}$ &  $\log \mathrm{L_{[OIII]5007, corr}}$ \\
  & [erg/s] & [erg/s] & [erg/s] & [erg/s] & [erg/s] & [erg/s] \\
\hline
J1652 & $44.01$ & $44.49$ & $43.79$ & $44.39$ & $44.71$ & $45.02$ \\
$1$ & $42.20$ & $42.67$ & $42.03$ & $42.63$ & $42.90$ & $43.36$ \\
$2$ & / & / & / & / & / & / \\
$3$ & $41.46$ & $41.93$ & $41.77$ & $42.37$ & $42.19$ & $42.65$ \\
$4$ & $41.65$ & $42.13$ & $41.75$ & $42.35$ & $42.07$ & $42.53$ \\
$5$ & $41.46$ & $41.94$ & $41.66$ & $42.27$ & $42.44$ & $42.90$ \\
$6$ & $41.35$ & $41.82$ & $41.82$ & $42.43$ & $42.16$ & $42.62$ \\
$7$ & $41.41^\star$ & $41.89^\star$ & / & / & $41.92^\star$ & $42.22^\star$ \\
$8$ & $40.32^\star$ & $40.79^\star$ & / & / & $41.29^\star$ & $41.60^\star$ \\
$9$ & $41.30^\star$ & $41.77^\star$ & / & / & $41.68^\star$ & $41.99^\star$ \\
$10$ & $42.56$ & $43.04$ & $42.71$ & $43.31$ & $43.03$ & $43.49$ \\
$11$ & $41.62$ & $42.09$ & $42.12$ & $42.73$ & $42.30$ & $42.76$ \\
$12$ & $42.27$ & $42.74$ & $42.50$ & $43.11$ & $42.78$ & $43.24$ \\
$13$ & $41.88$ & $42.36$ & $41.93$ & $42.53$ & $42.56$ & $43.02$ \\
$14$ & $42.08$ & $42.55$ & $42.38$ & $42.99$ & $42.81$ & $43.27$ \\
$15$ & $41.85$ & $42.32$ & $41.95$ & $42.55$ & $42.44$ & $42.90$ \\
$16$ & / & / & / & / & / & / \\
$17$ & $41.72$ & $42.20$ & $41.94$ & $42.54$ & $42.38$ & $42.84$ \\
\hline
\end{tabular}
\tablefoot{
{The columns show the observed fluxes, as well as dust-corrected ones using our fiducial assumption of $A_V = 0.7$ (Section \ref{sec:PC_members}). IDs 2 and 16 correspond to the passive galaxies identified via their 4000\AA\ break instead. \\
$^\star$ Measured from the zoomed-in JWST/NIRSpec data \citep{Wylezalek2022b}, as the NIRISS slitless spectra of the companion galaxies are contaminated by the bright quasar.}
}
\end{minipage}
\end{table*}

\section{Potential additional source}
\label{app:pot_extra}

\begin{figure*}
    \centering
    \includegraphics[width=0.62\linewidth, valign=m]{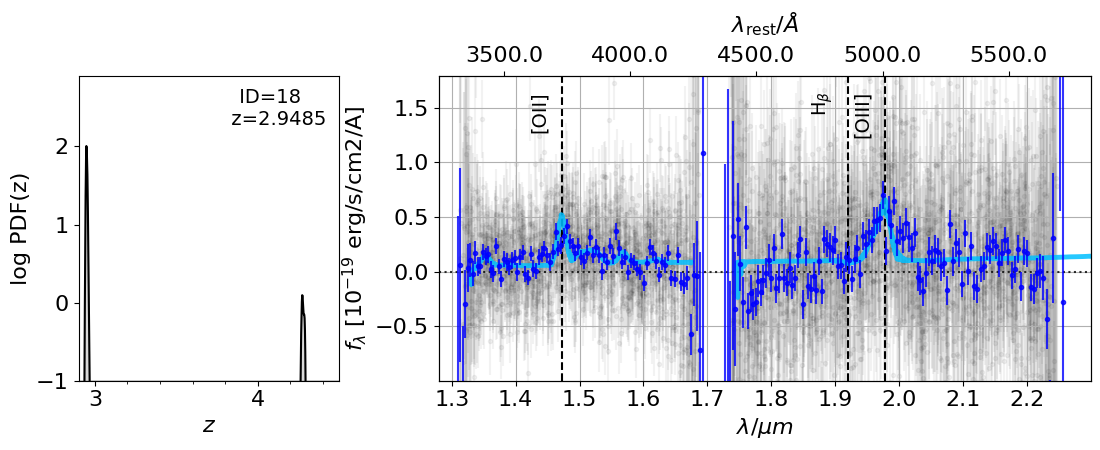}
    \includegraphics[width=0.23\linewidth, valign=m, trim={6cm 1.5cm 6cm 2cm}, clip]{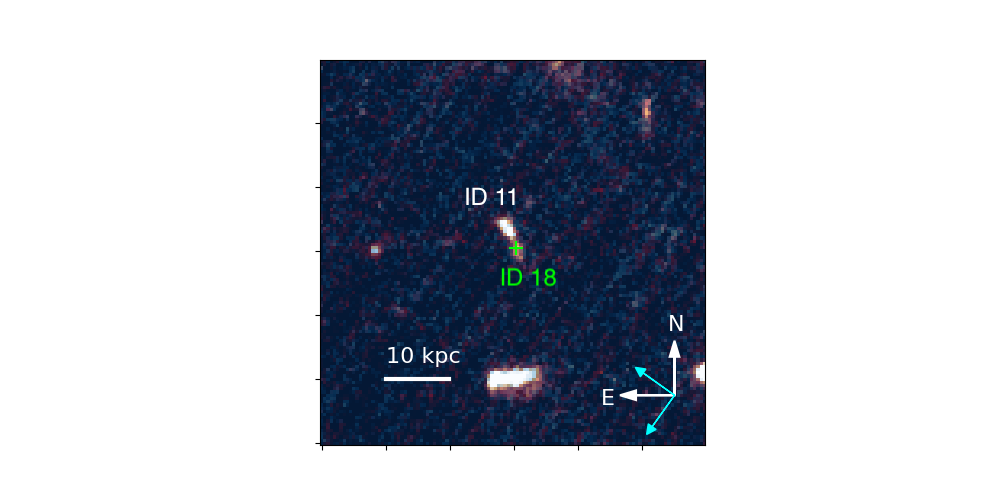}
    \includegraphics[width=0.47\linewidth]{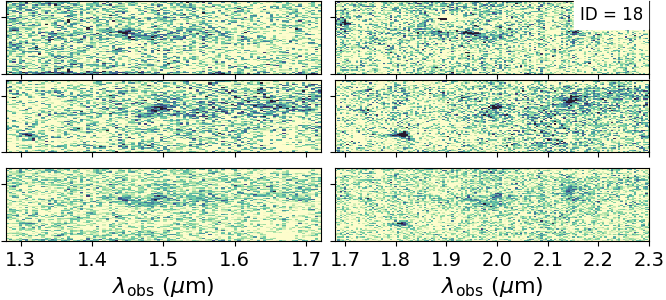}
    \caption{{Top}: Spectrum and recovered redshift probability distribution of a potential additional cluster member, which we do not include in the analysis. The right panel shows the HST WFC3 (blue) and ACS (red) photometry as a composite image, where the orientation of the grisms is indicated by cyan arrows. The source may be undergoing a merger with ID 11 in Fig \ref{fig:other_spectra}; however the posterior distribution in redshift has an alternative peak at higher redshift ($z = 4.23$). {Bottom}: 2D spectrum of the source, plotted analogously to Fig \ref{fig:2D_spectra}. The lines appear offset in the two grisms, which may be due to contamination by ID 11 in one grism. }
    \label{fig:spectrum_extra}
\end{figure*}

In Fig \ref{fig:spectrum_extra}, we present a potential additional cluster member that could be a companion to ID 12. While the best-fit redshift is 2.9485, there is an alternative, lower probability solution at $z=4.2$. Given the noisy character of the spectrum and the lack of unambiguous redshift determination, we do not include this source in our calculations.

\end{appendix}

\end{document}